\documentclass[sigconf]{acmart}

\renewcommand\footnotetextcopyrightpermission[1]{}

\newcommand{\etal}{\textit{et al. }}
\usepackage{url}
\usepackage[ruled,linesnumbered]{algorithm2e}
\usepackage{multirow}

\begin{document}
\fancyhead{}
\title{ConAML: Constrained Adversarial Machine Learning for Cyber-Physical Systems}

\author{Jiangnan Li}
\email{jli103@vols.utk.edu}
\affiliation{University of Tennessee, Knoxville}

\author{Yingyuan Yang}
\email{yyang260@uis.edu}
\affiliation{University of Illinois Springfield}

\author{Jinyuan Stella Sun}
\email{jysun@utk.edu}
\affiliation{University of Tennessee, Knoxville}

\author{Kevin Tomsovic}
\email{tomsovic@utk.edu}
\affiliation{University of Tennessee, Knoxville}

\author{Hairong Qi}
\email{hqi@utk.edu}
\affiliation{University of Tennessee, Knoxville}

\begin{abstract}

Recent research demonstrated that the superficially well-trained machine learning (ML) models are highly vulnerable to adversarial examples. As ML techniques are becoming a popular solution for cyber-physical systems (CPSs) applications in research literatures, the security of these applications is of concern. However, current studies on adversarial machine learning (AML) mainly focus on pure cyberspace domains. The risks the adversarial examples can bring to the CPS applications have not been well investigated. In particular, due to the distributed property of data sources and the inherent physical constraints imposed by CPSs, the widely-used threat models and the state-of-the-art AML algorithms in previous cyberspace research become infeasible.

We study the potential vulnerabilities of ML applied in CPSs by proposing Constrained Adversarial Machine Learning (ConAML), which generates adversarial examples that satisfy the intrinsic constraints of the physical systems. We first summarize the difference between AML in CPSs and AML in existing cyberspace systems and propose a general threat model for ConAML. We then design a best-effort search algorithm to iteratively generate adversarial examples with linear physical constraints. We evaluate our algorithms with simulations of two typical CPSs, the power grids and the water treatment system. The results show that our ConAML algorithms can effectively generate adversarial examples which significantly decrease the performance of the ML models even under practical constraints.

\end{abstract}


\maketitle

\section{Introduction}
\label{sec:intro}

Machine learning (ML) has shown promising performance in many real-world applications, such as image classification \cite{he2016deep}, speech recognition \cite{graves2013speech}, and malware detection \cite{yuan2014droid}. In recent years, motivated by the promotion of cutting-edge communication and computational technologies, there is a trend to adopt ML in various cyber-physical system (CPS) applications, such as data center thermal management \cite{Li:2011:TCF:2020408.2020611}, agriculture ecosystem management \cite{dabrowski2018state}, power grid attack detection \cite{ozay2015machine}, and industrial control system anomaly detection \cite{kravchik2018detecting}.

Recent research has demonstrated that the superficially well-trained ML models are highly vulnerable to adversarial examples \cite{szegedy2013intriguing, goodfellow2014explaining, rozsa2016adversarial, 7780651, kurakin2016adversarialscale, dong2018boosting, moosavi2017universal}. In particular, adversarial machine learning (AML) technologies enable attackers to deceive ML models with well-crafted adversarial examples by adding small perturbations to legitimate inputs. As CPSs have become synonymous to security-critical infrastructures such as the power grid, nuclear systems, avionics, and transportation systems, such vulnerabilities can be exploited leading to devastating consequences.

\begin{figure}[htbp]
\centerline{\includegraphics[width=0.85\linewidth]{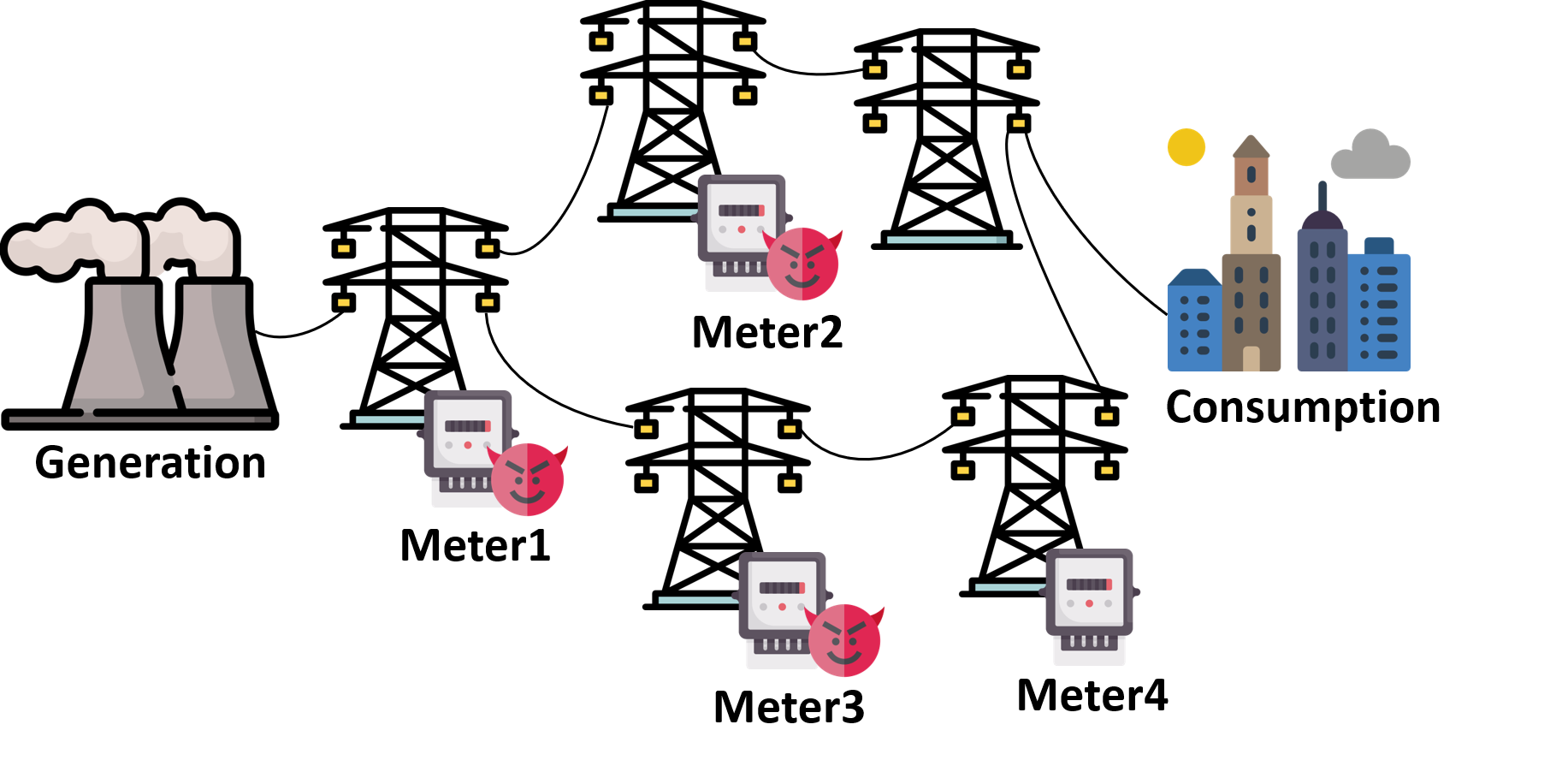}}
\caption{A CPS example (power grids).}
\label{fig:powerExa}
\end{figure}

AML research has received considerable attention in artificial intelligence (AI) communities and it mainly focuses on computational applications such as computer vision. However, it is not applicable to CPSs because the inherent properties of CPSs render the widely-used threat models and AML algorithms in previous research infeasible. The existing AML research makes common assumptions on the attacker's knowledge and the adversarial examples. In most AML research, the attacker is assumed to have full knowledge of the ML inputs and these features are assumed to be mutually independent. For example, in computer vision \cite{goodfellow2014explaining}, the attacker is assumed to know all the values of pixels of an image and there is no strict dependency among the pixels. However, this is not realistic for attacks targeting CPSs. CPSs are usually large and complex systems whose data sources are heterogeneous and geographically distributed. The attacker may compromise a subset of sensors and modify their measurement data. Generally, for the uncompromised data sources, the attacker cannot even know the measurements, let alone making modifications. Furthermore, for robustness and resilience reasons, CPSs usually employ redundant data sources and incorporate faulty data detection mechanisms. For example, in the power grid, redundant phasor measurement units (PMUs) are deployed in the field to measure frequency and phase angle, and residue-based bad data detection is employed to detect and recover from faulty data for state estimation \cite{wood2013power}. Therefore, the features of ML applications in CPS are not only dependent but also subject to the physical constraints of the system. A simple example of constraints is shown in Figure \ref{fig:powerExa}. All three meters are measuring the electric current (Ampere) data. If an attacker compromises \textit{Meter1}, \textit{Meter2}, and \textit{Meter3}, no matter what modification the attacker makes to the measurements, the compromised measurement of \textit{Meter1} should always be the sum of that of \textit{Meter2} and \textit{Meter3} due to Kirchhoff's laws. Otherwise, the crafted measurements will be detected by the bad data detection mechanism and obviously anomalous to the power system operators. In addition to distributed data sources and physical constraints, sensors in real-world CPSs are generally configured to collect data with a specific sampling rate. A valid adversarial attack needs to be finished within the CPS' sampling period.

The intrinsic properties of CPS pose stringent requirements for the adversarial attackers. The attacker is now required to overcome:
\begin{itemize}
\setlength\itemsep{0em}
\item \textbf{Knowledge constraint: } No access to the ML models and the measurement values of uncompromised sensors.

\item \textbf{Physical constraint: } The adversarial examples need to meet the physical constraints defined by the system.

\item \textbf{Time constraint: } Attacks needs to be completed within a sample period of the sensors.

\end{itemize}

to launch an effective attack that deceives the ML applications deployed in CPSs. However, in this paper, we show that the ML applications in CPSs are susceptible to handcrafted adversarial examples even though such systems naturally pose a greater barrier for the attacker.

In this paper, we propose constrained adversarial machine learning (\textbf{ConAML}), a general AML framework that incorporates the above constraints of CPSs. We firstly design a universal adversarial measurement algorithm to solve the knowledge constraint. After that, without loss of generality, we present a practical best-effort search algorithm to effectively generate adversarial examples under linear physical constraints which are one of the most common constraints in real-world CPS applications, such as power grids \cite{monticelli2012state} and water pipelines \cite{goh2016dataset}. Meanwhile, we set the maximum iteration number to control the time cost of the attack. We implement our algorithms with ML models used in two CPSs and mainly focus on neural networks due to its transferability. Our main contributions are summarized as follows:

\begin{itemize}

\item We highlight the potential vulnerability of deploying ML in CPSs, analyze the different requirements for AML applied in CPSs with regard to the general computational applications, and present a practical threat model for AML in CPSs. 

\item We formulate the mathematical model of ConAML by incorporating the physical constraints of the underlying system. 

\item We proposed ConAML, an AML framework that contains a series of AML algorithms to generate adversarial examples under the corresponding constraints.

\item We assess our algorithms with two typical CPSs, the power grids and water treatment system, where ML are intensively investigated for attack detection in the research literature \cite{ozay2015machine, yan2016detection, he2017real, james2018online, ayad2018detection, 8791598, wang2020detection, inoue2017anomaly, kravchik2018detecting, feng2019systematic, chen2018learning, feng2017deep, ahmed2018noise}. The evaluation results show that the adversarial examples generated by our algorithms can effectively bypass the ML-powered attack detection systems in the two CPSs.

\end{itemize}

Related research is discussed in Section \ref{sec:related}. We analyze the properties of AML in CPSs and give the mathematical definition and the threat model in Section \ref{sec:threat}. Section \ref{sec:design} presents the algorithm design. Section \ref{sec:evaluation} uses two CPSs as proofs of concept to carry our experiments. Discussions and future work are given in Section \ref{sec:future}. Section \ref{sec:conclusion} concludes the paper.

\section{Related Work}
\label{sec:related}

AML of deep neural network (DNN) was discovered by Szegedy \etal \cite{szegedy2013intriguing} in 2013. They found that a DNN used for image classification can be fooled by adding a hardly perceptible perturbation to the legitimate image. The same perturbation can cause a different DNN to misclassify the same image even when the DNN has a different structure and is trained with a different dataset, which is referred to as the transferability property of adversarial examples. In 2015, Goodfellow \etal \cite{goodfellow2014explaining} proposed the Fast Gradient Sign Method (FGSM), an efficient algorithm to generate adversarial examples. The Fast Gradient Value (FGV) method by Rozsa \etal \cite{rozsa2016adversarial} is a variant of FGSM and utilizes the raw gradient instead of the sign values. Moosavi-Dezfooli \etal presented \textit{DeepFool} to iteratively search for the closest distance between the original input and the decision boundary \cite{7780651}. Single-step attacks have better transferability but can be easily defended \cite{kurakin2016adversarialscale}. Therefore, multi-steps methods, such as iterative methods \cite{kurakin2016adversarialscale} and momentum-based methods \cite{dong2018boosting}, are presented. The above methods generate individual adversarial examples for each input. In 2017, Moosavi-Dezfooli \etal designed universal adversarial perturbations to generate perturbations regardless of the ML model inputs \cite{moosavi2017universal}. 

Research on AML applications continues growing rapidly. Sharif \etal launched adversarial attacks to a face-recognition system and achieved a notable result \cite{sharif2016accessorize}. Grossee \etal constructed adversarial attacks against Android malware detection models \cite{grosse2017adversarial}. In 2014, Laskov \etal developed a taxonomy for practical adversarial attacks based on the attackers' capability and launched evasion attacks to \textit{PDFRATE}, a real-world online machine learning system to detect malicious PDF malware \cite{laskov2014practical}. In 2018, Li \etal presented \textit{TEXTBUGGER}, a framework to generate adversarial text against deep learning-based text understanding (DLTU) systems and achieved state-of-the-art attack performance \cite{li2018textbugger}.


AML techniques that involve the physical domain are drawing more and more attention. Kurakin \etal presented that ML models are vulnerable to adversarial examples in physical world scenarios by feeding a phone camera captured adversarial image to an ImageNet classifier \cite{kurakin2016adversarial}. In 2016, Carlini \etal presented that well-crafted voice commands which are unintelligible to human listeners, can be interpreted as commands by voice controllable systems \cite{carlini2016hidden}. \cite{tian2018deeptest} and \cite{lu2017no} investigated the security of ML models used in autonomous driving cars. In 2018, \cite{ijcai2018-524} showed that an attacker can generate adversarial examples by modifying a portion of measurements in CPSs, and presented an anomaly detection model where each sensor's reading is predicted as a function of other sensors' readings. After that, Erba \etal also studied the AML in CPS and consider the physical constraints \cite{erba2019real}. They employed an autoencoder that is trained on normal system data to reconstruct the bad inputs to match the physical behavior. However, both \cite{ijcai2018-524} and \cite{erba2019real} allow the attacker to know all the measurements which may be impractical in real-world attacks. Meanwhile, the generated adversarial examples of \cite{erba2019real} may still violate the physical constraints.

More related work on adversarial attacks, including the adversarial example generation and applications, can be found in \cite{yuan2019adversarial}.

\section{System and Threat Model}
\label{sec:threat}
\subsection{ML-Assisted CPSs}

\begin{figure}[htbp]
\centerline{\includegraphics[width=0.70\linewidth]{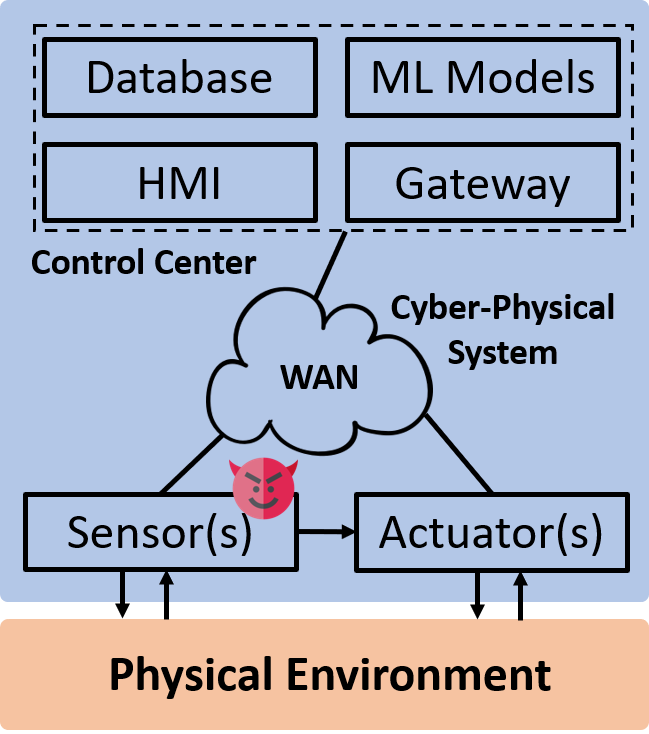}}
\caption{Machine learning-assisted CPS architecture.}
\label{fig:cps}
\end{figure}

Generally, a CPS can be simplified as a system that consists of four parts, namely sensors, actuators, the communication network, and the control center \cite{chen2018learning}, as shown in Figure \ref{fig:cps}. The sensors measure and quantify the data from the physical environment, and send the measurement data to the control center through the communication network. In practice, the raw measurement data will be filtered and processed by the gateway according to the error checking mechanism whose rules are defined by human experts based on the properties of the physical system. Measurement data that violates the physically defined rules will be removed. 

Similar to \cite{erba2019real}, we consider the scenario that the control center utilizes ML model(s) to make decisions (classification) based on the filtered measurement data from the gateway directly, and the features used to train the ML models are the measurements of sensors respectively. The target of the attacker will be deceiving the ML model(s) in CPSs to output wrong (classification) results without being detected by the gateway by adding perturbations to the measurements of the compromised sensors.

\subsection{Threat Model}
\label{subsec:threatmodel}

Adversarial attacks can be classified according to the attacker's capability and attack goals \cite{laskov2014practical, yuan2019adversarial, chakraborty2018adversarial}. In this work, we consider the integrity attack that the attacker generates adversarial perturbations to the ML inputs to deceive the ML model to make incorrect classification outputs.

There are several inherent properties of CPS that pose specific requirements for adversarial attacks. First, in CPS, ML models are usually placed in the control centers and other centralized locations which employ comprehensive and advanced security measures such as air-gapped networks. It is highly unlikely for the attacker to have access to the models and a black-box attack should be considered. Second, we assume that the attacker cannot access the training dataset for the same reason as above, but has access to an alternative dataset such as historical data that follows a similar distribution to train their models. It is possible for the attacker to obtain historical data in practice, for instance, temperature data for load forecasting, earthquake sensor data, flood water flow data, and traffic flow data, since these data are usually published or shared among multiple parties. 

To launch adversarial attacks, the attacker is assumed to compromise a certain number of sensors, and can freely eavesdrop and modify their measurement data. These sensors are deployed in the wild and their security is hard to guarantee. In real attack scenarios, this can be implemented by either directly compromising the sensors, such as device intrusion or attacking the communication network, such as man-in-the-middle attacks. However, due to the vastly distributed nature of sensors in CPS, it is only reasonable for the attacker to compromise a subset of the data sources but not all of them. For the uncompromised sensors, the attacker can neither know their measurement values nor make modifications. This constraint indicates that the attacker has limited knowledge of the ML inputs.

\begin{figure}[htbp]
\centerline{\includegraphics[width=0.65\linewidth]{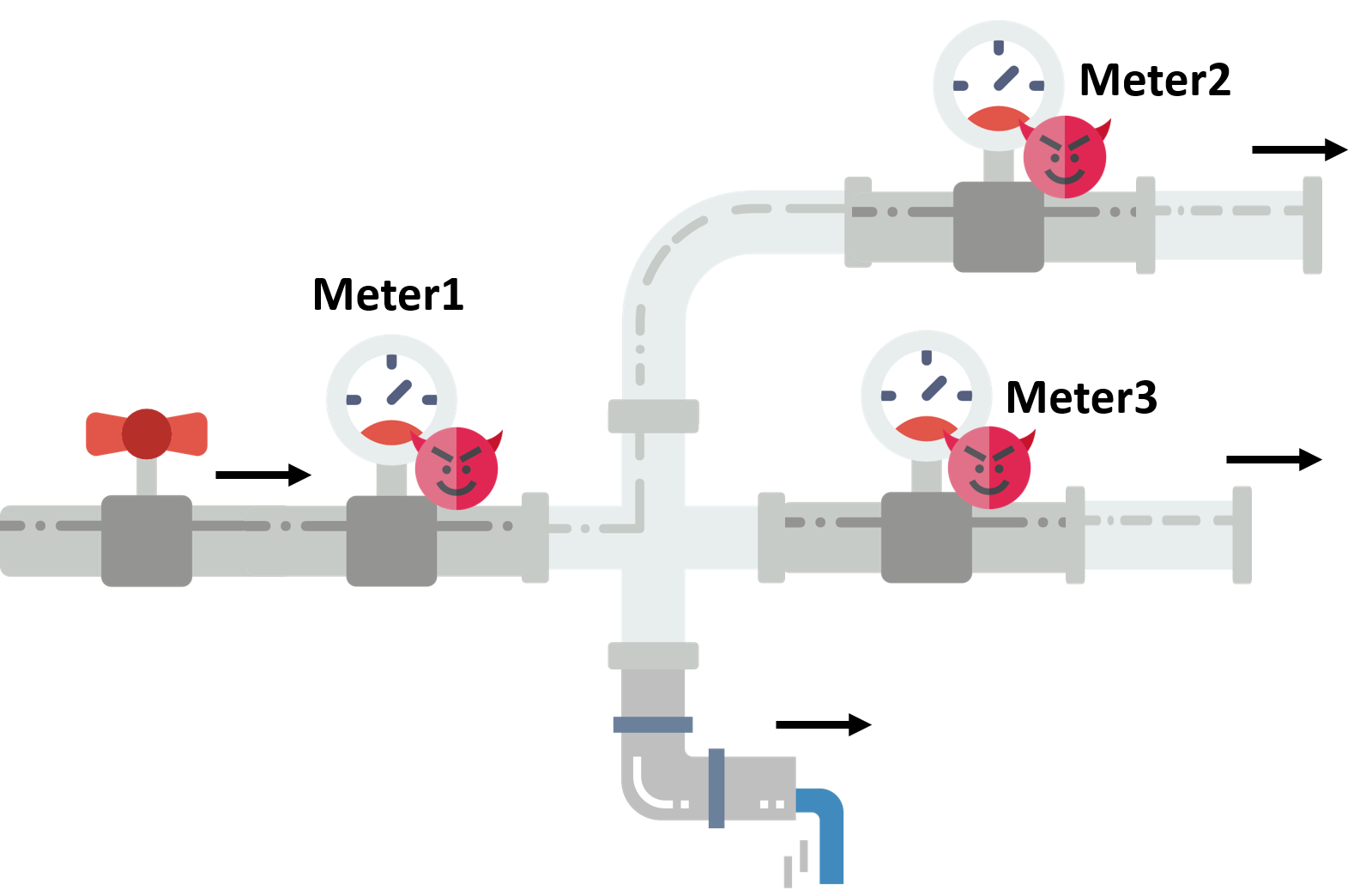}}
\caption{A CPS example (water pipelines).}
\label{fig:pipeline}
\end{figure}

Meanwhile, the attacker is further required to generate adversarial examples that meet the constraints imposed by the physical laws and system topology and evade any built-in detection mechanisms in the system. Specifically, since they are very common in real-world CPSs, we will mainly focus on linear constraints in this paper, including both linear equality constraints and linear inequality constraints. An example of the linear inequality constraint is shown in Figure \ref{fig:pipeline}. All the meters in Figure \ref{fig:pipeline} are measuring water flow which follows the arrows' direction. If an attacker wants to defraud the anomaly detection ML model of a water treatment system by modifying the meters' readings, the adversarial measurement of \textit{Meter1} should always be larger than the sum of \textit{Meter2} and \textit{Meter3} due to the physical structure of the pipelines. Otherwise, the poisoned inputs will be obviously anomalous to the victim (system operator) and detected automatically by the error checking mechanisms. In practice, many of the linear constraints can be explicitly abstracted by the attacker if she/he obtained enough measurement data by observing the compromised sensors. Meanwhile, the practical CPSs usually have built-in tolerance for noise and normal fluctuation in the measurements so that the approximately estimated constraints will still be effective for the adversarial attackers. Therefore, we assume that the attacker know the linear constraints among the compromised measurements. We discuss the nonlinear equality constraints at \textbf{Appendix \ref{app:nonlinear}}.

The real-world CPSs, such as the Supervisory Control and Data Acquisition (SCADA), will have a constant measurement sampling rate (frequency) configured for their sensors. The attacker who targets CPSs' ML applications is then required to generate a valid adversarial example within a measurement sampling period. 

We summarize the threat model as follows:

\begin{itemize}
\item We assume the attacker has no access to the system operator's trained model in the control center, including the hyper-parameters and the related dataset. However, the attacker has an alternative dataset as an approximation of the defender's (system operator's) training dataset to train his/her ML models.

\item The attacker can compromise a subset of sensors in the CPS and make modifications to their measurement data. However, the attack can neither know nor modify the measurements of uncompromised sensors. 

\item The attacker can know the linear constraints of the measurements imposed by the physical system.

\end{itemize}

\subsection{Physical Constraint Mathematical Representation}

In this subsection, we present the mathematical definition of the physical linear constraints of the ML inputs and represent the AML as a constrained optimization problem.

\subsubsection{\textbf{Notations}}

To simplify the mathematical representation, we will use $A_{B} = \left [ a_{b_{0}}, a_{b_{1}}, ..., a_{b_{n-1}} \right ]$ to denote a sampled vector of $A = \left [ a_{0}, a_{1}, ..., a_{m-1} \right ]$ according to $B$, where $B = \left [ b_{0}, b_{1}, ..., b_{n-1} \right ]$ is a vector of sampling index. For example, if $A = \left [a, b, c, d, e \right ]$ and $B = \left [ 0, 2, 4\right ]$, we have $A_{B} = \left [ a, c, e \right ]$.

We assume there are totally $d$ sensors in a CPS, and each sensor's measurement is a feature of the ML model $f_{\theta}$ in the control center. We use $S = \left [  s_{0}, s_{1}, ... , s_{d-1} \right ]^{T}$ and $M = \left [ m_{0}, m_{1}, ... , m_{d-1} \right ]^{T}$ to denote all the sensors and their measurements respectively. The attacker compromised $r$ sensors in the CPS and $C = \left [c_{0}, c_{1}, ..., c_{r-1} \right ]$ denotes the index vector of the compromised sensors. Obviously, we have $\left \| C \right \| = r $ and $ 0 < r \leq d$. Meanwhile, the uncompromised sensors' indexes are denoted as $U = \left [u_{0}, u_{1}, ..., u_{d-r-1} \right ]$ $ (\left \| U \right \| = d-r)$.

$\Delta =  \left [ \delta_{0}, \delta_{1}, ... , \delta_{d-1} \right ]^{T}$ is the adversarial perturbation to be added to $M$. However, the attacker can only inject $\Delta _{C} = \left [ \delta_{c_{0}}, \delta_{c_{1}}, ..., \delta_{c_{r-1}} \right ]^{T}$ to $M_{C}$ while $\Delta_{U} = 0$. The polluted adversarial measurements become $M_{C}^{\ast} = M_{C} + \Delta _{C}$, and $m_{c_{i}}^{\ast} = m_{c_{i}} + \delta_{c_{i}}$ $(0 \leq i \leq r - 1)$. Apparently, we have $\delta_{i} = \delta_{c_{j}}$ when $i = c_{j}$, $i \in C$, and $\delta_{i} = 0$ when $i \notin  C$. Similarly, the crafted adversarial example $M^{\ast} = \left [ m_{0}^{\ast}, m_{1}^{\ast}, ..., m_{d-1}^{\ast}\right ] = M + \Delta$ is fed into $f_{\theta}$. We have $m^{\ast}_{i} = m_{c_{j}}^{\ast}$ when $i = c_{j}$, $i \in C$ and $m^{\ast}_{i} = m_{i}$ when $i \notin C$. All the notations are summarized in Table \ref{table:notation}.

\begin{table}[htbp]
\caption{List of Notations}
\begin{center}
\begin{tabular}{|c|l|}
\hline
\textbf{Symbol} &  \multicolumn{1}{|c|}{\textbf{Description}}\\
\hline
$f_{\theta}$ & The trained model with hyperparameter $\theta$\\
\hline
$S$ & The vector of sensors \\
\hline
$M$ & The vector of measurements of $S$\\
\hline
$\Delta$ & The perturbations vector added to $M$ \\
\hline
$M^{\ast}$ & The sum of $\Delta$ and $M$. The vector of \\ 
&   compromised input\\
\hline
$C$ & The vector of the indexes of compromised \\
 &  sensors or measurements \\
\hline
$U$ & The vector of the indexes of uncompromised \\
 &  sensors or measurements \\
\hline
$Y$ & The original class of the measurement $M$\\
\hline
$\Phi$ & The linear constraint matrix\\
\hline
\end{tabular}
\end{center}
\label{table:notation}
\end{table}

\subsubsection{\textbf{Mathematical Presentation}}

For \textbf{linear equality} \textbf{constraints}, such as the current measurements (Amperes) of the three meters in Figure \ref{fig:powerExa}, we suppose there are $k$ constraints of the compromised measurements $M_{C}$ that the attacker needs to meet, and the $k$ constraints can be represented as follow:

\begin{equation}
\left\{
             \begin{array}{lr}
             \phi_{0,0}\cdot m_{c_{0}} + ... + \phi_{0,r-1}\cdot m_{c_{r-1}} = \phi_{0,r}\\
             \phi_{1,0}\cdot m_{c_{0}} + ... + \phi_{1,r-1}\cdot m_{c_{r-1}} = \phi_{1,r}\\
              ...\\
             \phi_{k-1,0}\cdot m_{c_{0}}  + ... + \phi_{k-1,r-1}\cdot m_{c_{r-1}} = \phi_{k-1,r}\\
             \end{array}
\right.
\end{equation}

The above constraints can be represented as (\ref{equ:linearEq}). We have $\Phi_{k \times r} = \left [ \Phi_{0}, \Phi_{1}, ..., \Phi_{k-1} \right ]^{T}$, where $\Phi_{i} = \left [ \phi_{i,0}, \phi_{i,1}, ..., \phi_{i,r-1} \right ]$ $(0\leq i\leq k-1)$, $\Phi_{i,j} = \phi_{i,j}$ $(0\leq i\leq k-1, 0\leq j\leq r-1)$ and $\tilde{\Phi} = \left [ \phi_{0,r}, \phi_{1,r}, ... , \phi_{k-1,r} \right ]^{T}$. 

\begin{equation}
\Phi_{k \times r}  M_{C} = \tilde{\Phi}
\label{equ:linearEq}
\end{equation}

The attacker generates the perturbation vector $\Delta _{C}$ and adds it to $M_{C}$ such that $f_\theta$ will predict the different output. Meanwhile, the crafted measurements $M^{\ast }_{C} = \Delta _{C} + M_{C}$ should also meet the constraints in (\ref{equ:linearEq}) to avoid being noticed by the system operator or detected by the error checking mechanism. 

Formally, the attacker who launches AML attacks needs to solve the following optimization problem:

\begin{subequations}
	\begin{align}
	\max_{\Delta _{C}} \;\;\;&L(f_{\theta}(M^{\ast}),Y)\\
	s.t. \;\;\; & M^{\ast }_{C} = M_{C} + \Delta _{C} \\ 
	&\Phi_{k \times r}  M_{C} = \tilde{\Phi}\\
	&\Phi_{k \times r}  M_{C}^{\ast} = \tilde{\Phi}\\
	&M^{\ast} = M + \Delta \\
	&\Delta_{U} = 0
	\end{align}
\label{eq:conOpt}
\end{subequations}

where $L$ is a loss function, and Y is the original class label of the input vector $M$.

In addition, the \textbf{linear inequality constraints} among the compromised measurements can be represented as equation (\ref{equ:linearIneq}), and the constrained optimization problem to be solved is also similar to (\ref{eq:conOpt}) but replacing (\ref{eq:conOpt}c) with $\Phi_{k \times r}  M_{C} \leq  \tilde{\Phi}$ and (\ref{eq:conOpt}d) with $\Phi_{k \times r}  M_{C}^{\ast} \leq \tilde{\Phi}$ respectively.

\begin{equation}
\Phi_{k \times r}  M_{C} \leq  \tilde{\Phi}
\label{equ:linearIneq}
\end{equation}

\section{Design of ConAML}
\label{sec:design}

The universal adversarial measurements algorithm is proposed in subsection \ref{subsec:univer} to solve the knowledge constraint of the attacker. Subsection \ref{subsec:linearAnay} and subsection \ref{subsec:linearIneGen} analyze the properties of physical linear equality constraints and linear inequality constraints in AML respectively and present the adversarial algorithms. We set the maximum numbers of searching step in different algorithms to control the attack's time cost.

\subsection{Universal Adversarial Measurements}
\label{subsec:univer}

\begin{algorithm}
\SetAlgoLined
\textbf{Input:} $f_{\theta}$, $MU$, $M_{C}$,  $\lambda$, $Y$, $MaxItera$\\
\textbf{Output:} $M^{\ast}$\\
\SetKwBlock{Begin}{function}{end function}
\Begin($\text{uniAdvMeasur} {(} f_{\theta}, MU, M_{C}, \lambda, Y, MaxItera {)}$)
{
  initialize $\Delta = 0$ \\
  build set $MUC = \left \{ M_{C|U_{0}}, M_{C|U_{1}}, ..., M_{C|U_{N}} \right \}$ \\
  set counter $cycNum = 0$ \\
  
  \While{$cycNum < MaxItera$}{
  set $flag$ to $0$ \\
  
  \For{$M_{C|U_{i}}$ in $MUC$}{
  $\Delta =$ \textbf{onePerturGenAlgorithm}$(\Delta, M_{C|U_{i}})$ \\
    \If{\textbf{sampleEva}$(f_{\theta}, Y, MUC, \Delta) <  \lambda$}{
    set $flag$ to $1$ \\
    \textbf{break}\\
  }
  }
  \If{$flag$ equals $1$}{
    \textbf{break}\\
  }
  $cycNum$++ \\
  }
  
  
  \Return{$M^{\ast} = M + \Delta$}
}
\caption{Universal Adv-Measur Algorithm}
\label{alg:univer}
\end{algorithm}

We first deal with the challenge of the attacker's limited knowledge on the uncompromised measurements $M_{U}$. This challenge is difficult to tackle since the complete measurement vector $M$ is needed to obtain the gradient values in many AML algorithms  \cite{goodfellow2014explaining, rozsa2016adversarial, 7780651, kurakin2016adversarialscale, moosavi2017universal}. In 2017, Moosavi-Dezfooli \etal proposed the universal adversarial perturbation scheme which generates image-agnostic adversarial perturbation \cite{moosavi2017universal}. The identical universal adversarial perturbation vector can cause different images to be misclassified by the state-of-the-art ML-based image classifiers with high probability. The basic philosophy of \cite{moosavi2017universal} is to iteratively and incrementally build a perturbation vector that can misclassify a set of images sampled from the whole dataset. 

\begin{algorithm}
\SetAlgoLined
\textbf{Input:} $f _{\theta}$, $Y$, $MUC$, $\Delta$ \\
\textbf{Output:} Classification Accuracy\\
\SetKwBlock{Begin}{function}{end function}
\Begin($\text{sampleEva} {(} f_{\theta}, Y, MUC, \Delta {)}$)
{
  add perturbation $\Delta$ to all vectors in $MUC$\\
  evaluate $MUC$ with $f_{\theta}$ and label $Y$\\
  return the classification accuracy of $f_{\theta}(MUC)$ \\
}
\caption{Sample Evaluation}
\label{alg:sampleEva}
\end{algorithm}

Inspired by their approach, we now present our universal adversarial measurements algorithm. We define an ordered set of $N$ sampled uncompromised measurements $MU = \left \{ M_{U_{0}}, M_{U_{1}},...,  M_{U_{N-1}} \right \}$, and use $M_{C|U_{i}}$ to denote the crafted measurement vector from $M_C$ and the sampled uncompromised measurement vector $M_{U_{i}}$. Here, $M_{C|U_{i}}$ is a crafted measurement vector with $\left \| M_{C|U_{i}} \right \| = d$. The uncompromised measurement vectors in $MU$ can be randomly selected from the attacker's alternative dataset.

Algorithm \ref{alg:univer} describes a high-level approach to generate adversarial perturbations regardless of uncompromised measurements. The algorithm first builds a set of crafted measurement vector $MUC$ based on $MU$ and $M_{C}$, and then starts an iteration over $MUC$. The iteration process is limited to $MaxItera$ times to control the maximum time cost. The purpose is to find a universal $\Delta$ that can cause a portion of the vectors in $MUC$ misclassified by $f_{\theta}$. The function \textbf{sampleEva} described in Algorithm \ref{alg:sampleEva} evaluates $MUC$ and $Y$ with the ML model $f_{\theta}$ and returns the classification accuracy. $\lambda \in \left(  0, 1\right ]$ is a constant chosen by the attacker to determine the attack's success rate in $MUC$ according to $\Delta$. During each searching iteration, algorithm \ref{alg:univer} builds and maintains the perturbation $\Delta$ increasingly using an adversarial perturbation generation algorithms, as shown by Line 10 in Algorithm \ref{alg:univer}. We will propose our methods to handle this problem in the next subsections.

\begin{figure}[htbp]
\centerline{\includegraphics[width=0.75\linewidth]{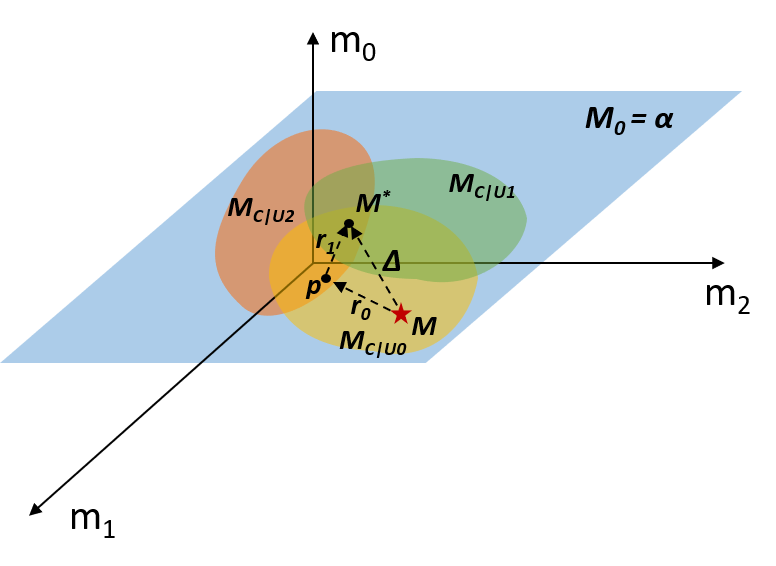}}
\caption{Iteration illustration.}
\label{fig:itelii}
\end{figure}

Figure \ref{fig:itelii} presents a simple illustration of the iteration process in Algorithm 1. We assume there are three sensors' measurements $M = \left [ m_{0}, m_{1}, m_{2} \right ]$ in a CPS and only one sensor's measurement $m_{0} = \alpha$ is compromised by the attacker. We set the the sample number $N=3$ and the yellow, green and orange shallow areas in the plane $M_{0} = \alpha$ represent the possible adversarial examples of the crafted measurement vector $M_{C|U_{0}}$, $M_{C|U_{1}}$, and $M_{C|U_{2}}$, respectively, where $U_{i}$ are randomly sampled measurements of uncompromised sensors ($m_{1}$ and $m_{2}$). The initial point $M$ (red $\bigstar$) iterates twice ($r_{0}$ and $r_{1}$) and finally reaches $M^{\ast}$ with the universal perturbation vector $\Delta$. Therefore, $M^{\ast}$ is a valid adversarial example for all $M_{C|U_{i}} (i \in \left \{ 0,1,2 \right \})$.

\noindent \textbf{Comparison of Methods:} Our approach is different from \cite{moosavi2017universal} in several aspects. First, the approach proposed in \cite{moosavi2017universal} has identical adversarial perturbations for different ML inputs while our approach actually generates distinct perturbations for each $M$. Second, the approach in \cite{moosavi2017universal} builds universal perturbations regardless of the real-time ML inputs. However, as the attacker has already compromised a portion of measurements, it is more effective to take advantage of the obtained knowledge. In other words, our perturbations are `\textbf{universal}' for $M_{U}$ but `\textbf{distinct}' for $M$. Finally, the intrinsic properties of CPSs require the attacker to generate a valid adversarial example within a sampling period while there is no enforced limitation of the iteration time in \cite{moosavi2017universal}.

\subsection{Linear Equality Constraints Analysis}
\label{subsec:linearAnay}

As shown in \cite{goodfellow2014explaining} and \cite{rozsa2016adversarial}, the fundamental philosophy of AML can be represented as (\ref{eq:amlfound}).

\begin{equation}
M^{\ast} = M + \Delta = M + \epsilon \nabla_{M}L(f_{\theta}(M), Y)
\label{eq:amlfound}
\end{equation}

However, directly following the gradient will not guarantee the adversarial examples meet the constraints in (\ref{equ:linearEq}) and (\ref{equ:linearIneq}). With the constraints imposed by the physical system, the attacker is no longer able to freely add perturbation to original input using the raw gradient of the input vector. In this subsection, we will analyze how the linear equality constraints will affect the way to generate perturbation and use a simple example for illustration. The proofs of all the theorems and corollaries can be found in \textbf{Appendix} \textbf{\ref{app:proof}}.

Under the threat model proposed in Section \ref{subsec:threatmodel}, the constraint of (\ref{eq:conOpt}c) is always met due to the properties of the physical systems. We then consider the constraint (\ref{eq:conOpt}d).

\begin{theorem}
The sufficient and necessary condition to meet constraint (\ref{eq:conOpt}d) is $\Phi_{k \times r} \Delta _{C} = 0$.
\label{theorem:condition}
\end{theorem}

From Theorem \ref{theorem:condition} we can also derive a very useful corollary, as shown below.

\begin{corollary}
If $\Delta _{C_{0}}$, $\Delta _{C_{1}}$, ..., $\Delta _{C_{n}}$ are valid perturbation vectors that follow the constraints, then we have $\Delta_{C'} = \sum_{i=0}^{n} a_{i} \cdot \Delta_{C_{i}}$ is also a valid perturbation for the constraint $\Phi_{k \times r}$.
\label{corollary:combine}
\end{corollary}

Theorem \ref{theorem:condition} indicates that the perturbation vector to be added to the original measurements must be a solution of the homogeneous linear equations $\Phi_{k \times r} X = 0$. However, is this condition always met? 

\begin{theorem}
In practical scenarios, the attacker can always find a valid solution (perturbation) that meets the linear equality constraints imposed by the physical systems. 
\label{theorem:alwaysMet}
\end{theorem}

\begin{figure}[htbp]
\centerline{\includegraphics[width=0.75\linewidth]{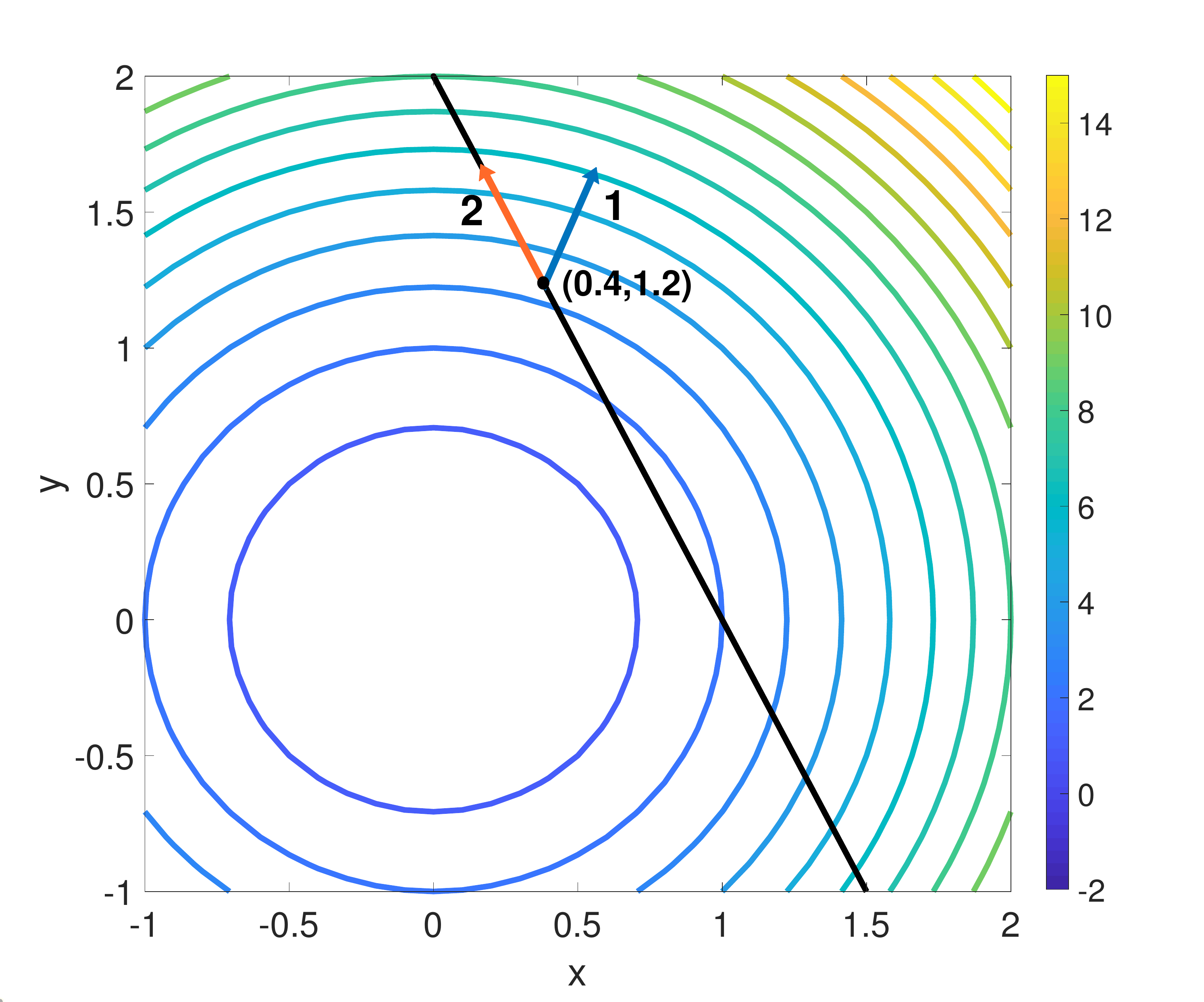}}
\caption{Linear equality constraint illustration. We consider a simple ML model $f$ that only has two dimensions inputs $(x,y)$ with a loss function $L(f_\theta(M), Y) = z(x,y) = 2x^{2} + 2y^{2}$. Meanwhile, we suppose the input measurements $x$ and $y$ need to meet the linear constraints $2x + y = 2$ and the current measurement vector $M = (0.4, 1.2)$.}
\label{fig:linearDemo}
\end{figure}

We utilize a simplified example to illustrate how the constraints will affect the generation of perturbations, as shown in Figure \ref{fig:linearDemo}. According to \ref{eq:amlfound}, measurement $M$ should move a small step (perturbation) to the gradient direction (direction 1 in Figure \ref{fig:linearDemo}) to increase the loss most rapidly. However, as shown by the contour lines in Figure \ref{fig:linearDemo}, the measurement $M$ is always forced to be on the straight line $y = 2 - 2x$ (2-dimension), which is the projection of the intersection of the two surfaces $z(x,y) = 2x^{2} + 2y^{2}$ and $2x + y = 2$ (3-dimension). Accordingly, instead of following the raw gradient, $M$ should move forward to direction 2 to increase the loss. Therefore, although at a relatively slow rate, it is still possible for the attacker to increase the loss under the constraints. 

\subsection{Adversarial Example Generation under Linear Equality Constraint }
\label{subsec:linearGen}

The common method of solving optimization problems using gradient descent under constraints is projected gradient descent (PGD). However, since neural networks are generally not considered as convex functions \cite{ChoromanskaHMAL14}, PGD cannot be used to generate adversarial examples directly. We propose the design of a simple but effective search algorithm to generate the adversarial examples under physical linear equality constraints.

\begin{algorithm}
\SetAlgoLined
\textbf{Input:} $\Delta$, $f _{\theta}$, $C$, $M$, $step$, $size$, $\Phi$, $Y$\\
\textbf{Output:} $v$\\

\SetKwBlock{Begin}{function}{end function}
\Begin($\text{genEqPer} {(} \Delta, f_{\theta}, C, M, step, size, \Phi, Y {)}$)
{
  initialize $v = \Delta$ \\
  initialize $stepNum = 0$ \\
  \While{$stepNum \leq step - 1$ }{
  \If{$f'_{\theta'}(M + v)$ doesn't equals $Y$}{
    \Return{$v$}
  }
  $r = \textbf{eqOneStep}(f_{\theta}, C,M + v, size, \Phi, Y)$\\
  update $v = v + r$ \\
  $stepNum = stepNum + 1$ \\
 }
 \Return{$v$}
}
\caption{Best-Effort Search (Linear Equality)}
\label{alg:searchEq}
\end{algorithm}

As discussed in subsection \ref{subsec:linearAnay}, the perturbation $\Delta _{C}$ needs to be a solution of $\Phi_{k \times r} X = 0$. We use $n = Rank(\Phi_{k \times r})$ to denote the rank of the matrix $\Phi_{k \times r}$, where $0 < n < r$. It is obvious that the solution set of homogeneous linear equation $\Phi_{k \times r} X = 0$ will have $r-n$ basic solution vectors. We use $I = \left [ i_{0}, i_{1}, ..., i_{r-n-1}\right ]^{T}$ to denote the index of independent variables in the solution set, $D = \left [ d_{0}, d_{1}, ..., d_{n-1}\right ]^{T}$ to denote the index of corresponding dependent variables, and $B_{n\times(r-n)}$ to denote the linear dependency matrix of $X_{I}$ and $X_{D}$. Clearly, we have $X_{D_{n\times1}} = B_{n\times(r-n)} X_{I_{(r-n)\times 1}}$. For convenience, we will use $\left [ I, D, B \right ] = \textbf{dependency}(\Phi_{k \times r})$ to describe the process of getting $I$, $D$, $B$ from matrix $\Phi_{k \times r}$.

\begin{algorithm}
\SetAlgoLined
\textbf{Input:} $f _{\theta}$, $C$, $M$, $size$, $\Phi$, $Y$\\
\textbf{Output:} $r$\\

\SetKwBlock{Begin}{function}{end function}
\Begin($\text{eqOneStep} {(} f_{\theta}, C, M, size, \Phi_{k \times r}, Y {)}$)
{
  calculate gradient vector $ G = \nabla_{M} L(f_{\theta}(M), Y)$\\
  set all elements of $G_U$ in $G$ to zero \\
  \textbf{define} $G' = G_{C}$ \\
  obtain tuple $\left [ I, D, B\right ] = \textbf{dependency}(\Phi_{k \times r})$\\
  update $G'_{D} = B G'_{I}$ in $G'$\\
  $\epsilon = size / \textbf{max}(\textbf{abs}(G'))$ \\
  \Return{$r = \epsilon G$}
}
\caption{One Step Attack Constraint $\Delta_{C}$}
\label{alg:oneTimeEq}
\end{algorithm}

As shown in Algorithm \ref{alg:searchEq}, the function \textbf{genEqPer} takes $\Delta$ as an input and outputs a valid perturbation $v$ for $M$. Algorithm \ref{alg:searchEq} keeps executing \textbf{eqOneStep} for multiple times defined by $step$ to generate a valid $v$ increasingly. Function \textbf{eqOneStep} performs a single-step attack for the input vector and returns a one-step perturbation $r$ that matches the constraints defined by $\Phi$, which is shown in Algorithm \ref{alg:oneTimeEq}. Due to Corollary \ref{corollary:combine}, $\Delta$ and $v$ will also follow the constraints. To decrease the iteration time, similar to \cite{7780651}, the algorithm will return the crafted adversarial examples immediately as long as $f'_{\theta'}$ misclassifies the input measurement vector $M + v$, as shown by Line 7 in Algorithm 3.

The philosophy of function \textbf{eqOneStep} in algorithm \ref{alg:oneTimeEq} is very straightforward. From the constraint Matrix $\Phi$, we can get the independent variables $I$, dependent variables $D$ and the dependency matrix $B$ between them. We will simply keep the gradient values of $I$ and use them to compute the corresponding values of $D$ (Line 8) so that the final output perturbation $r$ will follow $\Phi$. The constant $size$ defines the largest modification of a specific measurement value in one iteration to control the search speed.

\subsection{Adversarial Example Generation under Linear Inequality Constraint}
\label{subsec:linearIneGen}

\begin{algorithm}
\SetAlgoLined
\textbf{Input:} $f' _{\theta'}$, $U$, $M$, $size$, $Y$\\
\textbf{Output:} $r$\\
\SetKwBlock{Begin}{function}{end function}
\Begin($\text{freeStep} {(} f'_{\theta'}, U, M, size, Y {)}$)
{
  calculate gradient vector $ G = \nabla_{M} L(f'_{\theta'}(M), Y)$\\
  set elements in $G_{U}$ to zero\\
  $\epsilon = size / \textbf{max}(\textbf{abs}(G))$ \\
  \Return{$r = \epsilon G$}
}
\caption{Non-Constraint Perturbation.}
\label{alg:freeStep}
\end{algorithm}

Linear inequality constraints are very common in real-world CPS applications, like the water flow constraints in Figure \ref{fig:pipeline}. Due to measurement noise, real-world systems usually tolerate distinctions between measurements and expectation values as long as the distinctions are smaller than predefined thresholds, which also brings inequality constraints to data. Meanwhile, a linear equality constraint can be represented by two linear inequality constraints. As shown in equation (\ref{equ:linearIneq}), linear inequality constraints define the valid measurement subspace whose boundary hyper-planes are defined by equation (\ref{equ:linearEq}). In general, the search process under linear inequality constraints can be categorized into two situations. The first situation is when a point (measurement vector) is in the subspace and meets all constraints, while the second situation happens when the point reaches boundaries.

\begin{algorithm}
\SetAlgoLined
\textbf{Input:} $\Delta$, $f'_{\theta'}$, $C$, $U$, $M$, $step$, $size$, $\Phi$, $\tilde{\Phi}$, $Y$\\
\textbf{Output:} $v$\\
\SetKwBlock{Begin}{function}{end function}
\Begin($\text{genIqPer} {(} \Delta, f'_{\theta'}, C, U, M, step, size, \Phi, \tilde{\Phi}, Y {)}$)
{
  initialize $pioneer$ = $\Delta$, $valid$ = $pioneer$ \\
  initialize $stepNum$ = $0$ \\
  initialize $V$ as empty // violated constrain index \\

  \While{$stepNum \leq step - 1$ }{
  \If{ $f'_{\theta'}(M + valid)$ doesn't equals $Y$}{
    \textbf{break}
  }
  $chkRst = \textbf{chkIq}(\Phi, \tilde{\Phi}, M + pioneer, C)$ \\
  \eIf{ $chkRst$ is empty} { 
    $valid$ = $pioneer$ \\
    $r =  \textbf{freeStep}(f'_{\theta'}, U, M + valid, size, Y)$ \\
    $pioneer = valid + r$ \\
    \textbf{reset} $V$ to empty \\
  }{
    extend $V$ with $chkRst$ \\
    \textbf{define} $ \Phi' = \Phi_{V}$ // real-time constraints\\ 
    $ r = \textbf{eqOneStep}(f'_{\theta'}, C, M + valid, size, \Phi', Y)$ \\
    $pioneer = valid + r$ \\
  }
  
  $stepNum = stepNum + 1$ \\
 }
 \Return{$v = valid$}
}
\caption{Best-Effort Search (Linear Inequality)}
\label{alg:searchIneq}
\end{algorithm}

Due to the property of physical systems, the original point $M$ will naturally meet all the constraints. As shown in Algorithm \ref{alg:searchIneq}, to increase the loss, the original point will first try to move a step following the gradient direction through the function \textbf{freeStep} defined in Algorithm \ref{alg:freeStep}. Algorithm \ref{alg:freeStep} is very similar to the FGM algorithm \cite{rozsa2016adversarial} but no perturbation is added to $M_{U}$, namely $r_{U} = 0$, which is similar to the saliency map function used in \cite{papernot2016limitations}. After that, the new point $M'$ is checked with equation (\ref{equ:linearIneq}) to find if all inequality constraints are met. If all constraints were met, the moved step was valid and we can update $M = M'$. If $M'$ violates some constraints in $\Phi$, we will take all the violated constraints and make a real-time constraint matrix $\Phi_{V}$, where $V$ is the index vector of violated constraints. We now convert the inequality constraint problem to the equality constraint problem with the new constraint matrix $\Phi_{V}$ and the original point $M$. $M$ will then try to take a step using the \textbf{eqOneStep} function described in Algorithm \ref{alg:oneTimeEq} with the new constraint matrix $\Phi_{V}$. Again, we check whether the new reached point meets all the constraints. If there are still violated constraints, we extend $V$ with the new violated constraints. The search process repeats until reaching a valid $M'$ that meets all the constraints. For simplicity, we will use $chkRst = \textbf{chkIq}(\Phi, \tilde{\Phi}, M', C)$ to denote the checking process of a single search in one step movement, where $chkRst$ is the index vector of the violated constraints in the search.

\begin{figure}[htbp]
\centerline{\includegraphics[width=0.75\linewidth]{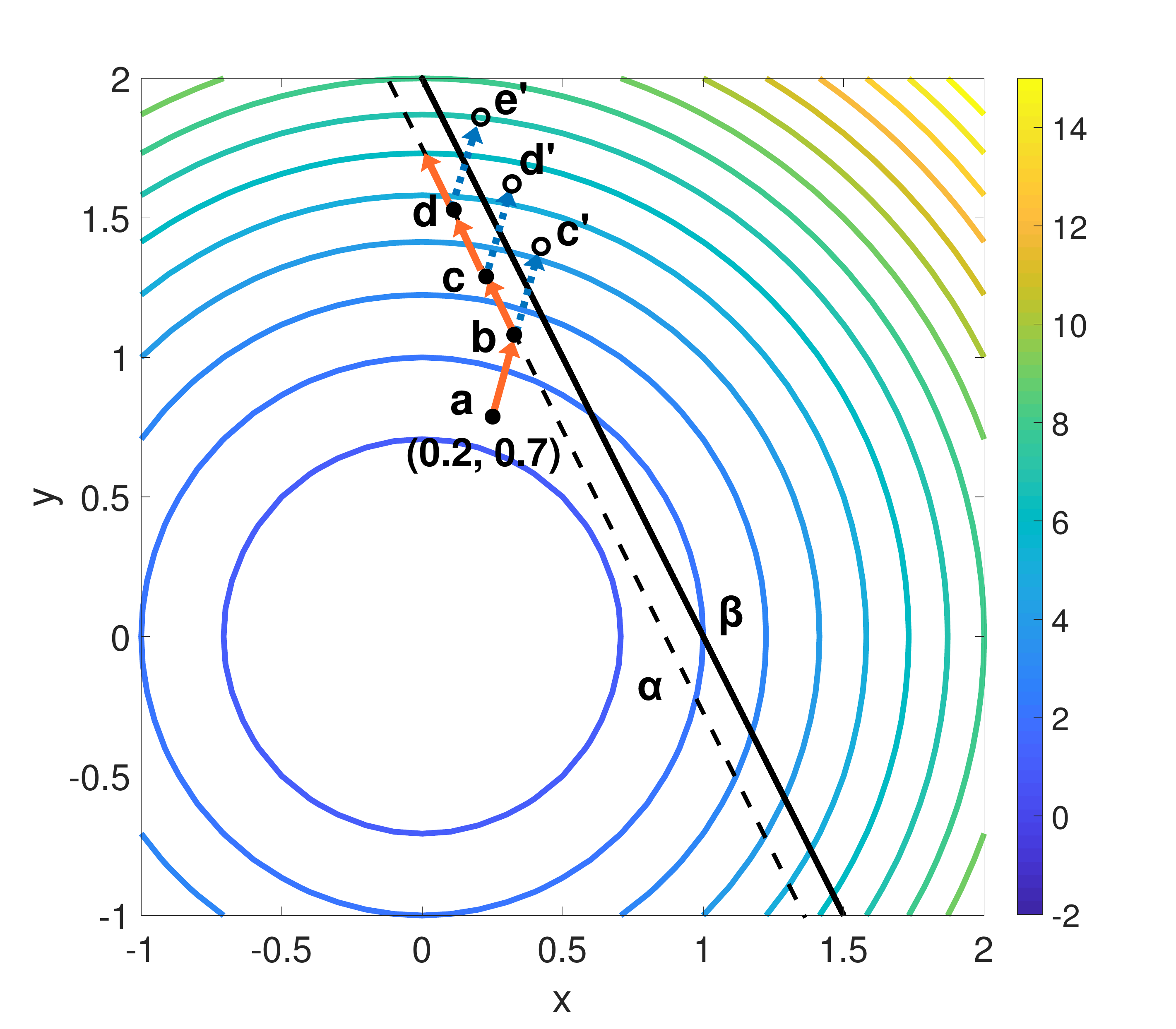}}
\caption{Best-Effort Search (linear inequality). We have the loss function $L(f_\theta(M), Y) = 2x^{2} + 2y^{2}$ with inequality constraints $y \leq 2 - 2x$.} 
\label{fig:ineqDemo}
\end{figure}

Similar to Figure \ref{fig:linearDemo}, a simple example is shown in Figure \ref{fig:ineqDemo}. To increase the loss, the initial point $a$ will take a small step following the gradient direction and reach point $b$. Since $b$ meets the constraints, it is a valid point. After that, $b$ will move a step following the gradient direction and reach point $c'$. However, point $c'$ violates the constraint $\beta$ and the movement is not valid. As we have point $b$ is valid, we construct a linear equality constraint problem with constraint $\alpha$ which is parallel to $\beta$. With constraint $\alpha$, point $b$ will move a step to point $c$ which is also a valid point. Point $c$ then repeats the search process and increases the loss gradually. The real-time equality constraint is only used once. When a new valid point is reached, it empties the previous equality constraints and tries the gradient direction first.


\section{Experimental Evaluation}
\label{sec:evaluation}

We evaluate our ConAML frameworks with two CPS study cases. The first study case is the ML-based false data injection attack (FDIA) detection in the power grids to show the impact of physical linear equality constraints, and the second case is the deep learning-based anomaly detection in the water treatment system to demonstrate linear inequality constraints. 

\noindent \textbf{Scenarios:} For each CPS study, we consider four attack scenarios regarded to the different knowledge constraints, as summarized in Table \ref{table:scenario}. The black-box scenario is the knowledge constraint we presented in our threat model in subsection \ref{subsec:threatmodel} and is the most practical scenario for the attacker. We consider different scenarios to show the impact of different constraints and to study the robustness of ML models in CPSs under different circumstances. For white-box and gray-box1 scenarios, the attacker won't execute Algorithm \ref{alg:univer} since the measurements of $M_{U}$ are given and there is no knowledge constraint. We note that attackers under all the scenarios in Table \ref{table:scenario} can only modify $M_{C}$ and should also follow the physical and time constraints.

\begin{table}[htbp]
\caption{Attack Scenarios}
\begin{center}
\begin{tabular}{|c|c|}
\hline
\textbf{Scenario} & \textbf{Constraint}\\
\hline
white-box & know both $M_{C}$ and $M_{U}$, has access to $f_{\theta}$\\
\hline
gray-box1 & know both $M_{C}$ and $M_{U}$, no access to $f_{\theta}$\\
\hline
gray-box2 & only know $M_{C}$, has access to $f_{\theta}$\\
\hline
black-box & only know $M_{C}$, no access to $f_{\theta}$\\
\hline
\end{tabular}
\end{center}
\label{table:scenario}
\end{table}

\noindent \textbf{Baselines:} We set two evaluation baselines in addition to the above scenarios. The first baseline is a supreme white-box attacker who has full access to the CPS ML model without considering any constraints and utilizes a state-of-the-art AML algorithm \cite{rozsa2016adversarial} to generate adversarial examples. We compare the performance of the ConAML framework with the supreme attack to demonstrate the impact of the constraints. The second baseline the autoencoder generator proposed in \cite{erba2019real} by Erba \etal in 2019. In \cite{erba2019real}, an autoencoder is trained with the normal CPS measurement data and is expected to learn the physical constraints of measurements. The adversarial examples are fed to the autoencoder to be transferred into examples that meet the physical constraints. However, \cite{erba2019real} allows the attacker to know complete $M$ (same as the gray-box1 scenario in Table \ref{table:scenario}), which is less practical compared with our threat model. Meanwhile, since the patterns are learned by neural networks (autoencoder), the transferred measurements may not meet the linear constraints strictly. In our experiment, we show that the generated examples from the autoencoder may still violate the physical constraints.

\noindent \textbf{Metrics:} The evaluation metrics of ConAML can be different according to the attack purpose and the CPS properties. We set three metrics to evaluate the attack performance in this study. The first metric is detection accuracy of the defender's model under attack and a lower detection accuracy indicates a better attack performance. The second metric is the magnitude of the noise injected to the legitimate measurement. The attacker needs the adversarial examples to bypass the detection while maintaining their malicious behavior. A small bad noise will violate the attack's original intention even it can bypass the detection. We select the $L_{2}$-Norm of the valid noise vector as the second metric to compare the magnitude of the malicious injected data. Finally, as the attack needs to be finished within a sampling period of the CPS, we will compare the time cost of the adversarial example generation.

\subsection{Case Study: State Estimation in Power Grids}
\subsubsection{Background: State Estimation and FDIA}

State estimation is a backbone of various crucial applications in power system control that has been enabled by large scale sensing and communication technologies, such as SCADA. It is used to estimate the state of each bus, such as voltage angles and magnitudes, in the power grid through analyzing other measurements. A DC model of state estimation can be represented as (\ref{eq:dcModel1}), where $\textbf{x}$ is the state, $\textbf{z}$ is the measurement, and \textbf{$\textbf{H}_{m \times n}$} is a matrix that determined by the topology, physical parameters and configurations of the power grid. 

\begin{equation}
\textbf{z} = \textbf{H}\textbf{x} + \textbf{e}
\label{eq:dcModel1}
\end{equation}

Due to possible meter instability and cyber attacks, bad measurements $\textbf{e}$ may be introduced to $\textbf{z}$. To solve this, the power system employs a residual-based detection scheme to remove the error measurements \cite{monticelli2012state}. The residual-based detection involves non-linear computation ($L_{2}$-Norm), however, research has shown that a false measurement vector following linear equality constraints can be used to pollute the normal measurements without being detected. In 2009, Liu \etal proposed the false data injection attack (FDIA) that can bypass the residual-based detection scheme and finally pollute the result of state estimation \cite{liu2009false}. In particular, if the attacker knows $\textbf{H}$, she/he could construct a faulty vector $\textbf{a}$ that meets the linear constraint $\textbf{Ba} = \textbf{0}$, where $\textbf{B} = \textbf{H}(\textbf{H}^{T}\textbf{H})^{-1}\textbf{H}^{T} - \textbf{I}$, and the crafted faulty measurements $\textbf{z} + \textbf{a}$ will not be detected by the system. A detailed introduction of state estimation, residual-based error detection, and FDIA can be found in \textbf{Appendix \ref{app:fdia}}

Many detection and mitigation schemes to defend FDIA are proposed, including strategical measurement protection \cite{bi2011defending} and PMU-based protection \cite{yang2017optimal}. In recent years, detection based on ML, especially neural networks, become popular in the literature \cite{ozay2015machine, yan2016detection, he2017real, james2018online, ayad2018detection, 8791598, wang2020detection}. The ML-based detection does not require extra hardware equipment and achieve the state-of-the-art detection performance. However, in this section, we will demonstrate that the attacker can construct an adversarial false measurement vector $ \textbf{z}_{adv}$ that can bypass both the residual-based detection and the ML-based detection. The ML models in previous research are trained to distinguish normal measurement $\textbf{z}$ and poisoned measurement $\textbf{z} + \textbf{a}$.  Our ConAML algorithms allow the attacker to generate an adversarial perturbation $\textbf{v}$ that meets the constraint $\textbf{Bv} = \textbf{0}$ for his/her original false measurement $\textbf{z} + \textbf{a}$ and obtain a new adversarial false measurement vector $ \textbf{z}_{adv} = \textbf{z} + \textbf{a} + \textbf{v}$ that will be classified as normal measurements by the ML-based FDIA detection models. The matrix $\textbf{B}$ then acts as the constraint matrix $\Phi$ defined in equation (\ref{eq:conOpt}). Meanwhile, $\textbf{z}_{adv}$ can naturally bypass the traditional residual-based detection approach since the total injected false vector $\textbf{a} + \textbf{v}$ meets the constraint $\textbf{B}(\textbf{a} + \textbf{v}) = \textbf{B}\textbf{a} + \textbf{B}\textbf{v} = \textbf{0}$. Our experiment in the next subsection will show that our ConAML algorithms can significantly decrease the detection accuracy of the ML-based detection schemes.

\subsubsection{Experiment Design and Evaluation}

We select the IEEE standard 10-machine 39-bus system as the power grid system as it is one of the benchmark systems in related research \cite{8791598,liu2016masking}. The features used for ML model training are the power flow (Ampere) measurements of each branches. The system has 46 branches so that there there will be 46 features for the ML models.  

The goal of the attacker is to implement a false-negative attack that makes $\textbf{z}_{adv}$ bypass the detections. We utilize the MATPOWER \cite{zimmerman2010matpower} library to derive the $\textbf{H}$ matrix and simulate related datasets. We simulate two training datasets for the system operator and the attacker to train their ML models. Through tuning the parameters, the overall detection accuracy of the defender's model $f_{\theta}$ is 98.3\% and the attacker's model $f'_{\theta'}$ is 97.5\%. After that, we assume there are 10, 13, and 15 measurements being compromised by the attacker and simulate the corresponding test datasets that only contain the false measurements. A more detailed description of the experiment can be found in \textbf{Appendix \ref{app:powerExper}}.

\begin{table}[htbp]
\caption{Evaluation Result Summary}
\begin{center}
\begin{tabular}{|c|c|c|c|c|}
\hline
\textbf{Attack} & \textbf{Case} & \textbf{Accu} & \textbf{$L_{2}$-Norm} & \textbf{Time (ms)}\\
\hline
\multirow{3}*{Supreme } & 10 & 0\% & 4077.43  & 5.8 \\
\cline{2-5}
  ~  & 13 & 0\% & 8403.84 & 12.9 \\
\cline{2-5}
 ~ & 15 & 0\% & 7979.26 & 6.8 \\
 \hline
 \multirow{3}*{Erba \cite{erba2019real}} & 10 & 0\% & 1049.52 & 5.7\\
\cline{2-5}
  ~  & 13 & 0\% & 1164.71 & 5.84\\
\cline{2-5}
 ~ & 15 & 0\% & 1578.87 & 5.94 \\
 \hline
 \multirow{3}*{white-box} & 10 & 0\% & 2527.8 & 42 \\
\cline{2-5}
  ~  & 13 & 0\% & 4984.03 & 96.8\\
\cline{2-5}
 ~ & 15 & 0\% & 7029.26 & 52.9 \\
 \hline
 \multirow{3}*{gray-box1} & 10 & 21.1\% & 2404.76 & 34.2\\
\cline{2-5}
  ~  & 13 & 48.9\% & 5356.09 & 87.1\\
\cline{2-5}
 ~ & 15 & 30.0\% & 9133.15 & 7.96 \\
 \hline
 \multirow{3}*{gray-box2} & 10 & 0\% & 2247.21 & 400.25\\
\cline{2-5}
  ~  & 13 & 5.4\% & 4882.95 & 222.4\\
\cline{2-5}
 ~ & 15 & 8.1\% & 6610.6 & 126.9\\
 \hline
  \multirow{3}*{black-box} & 10 & 14.4\% & 1843.2 & 131.9 \\
\cline{2-5}
  ~  & 13 & 4.3\% & 4786.72 & 209.6\\
\cline{2-5}
 ~ & 15 & 28.1\% & 9079.02 & 163.3 \\
 \hline
\end{tabular}
\end{center}
\label{table:powerEva}
\end{table}

Table \ref{table:powerEva} summarizes the detection performance of $f_{\theta}$ under different adversarial attacks generated by our ConAML algorithms. From the table, we can learn that the ConAML attacks can effectively decrease the detection accuracy of the ML models used for FDIA detection and inject considerable bad data to the state estimation, even under black-box scenario. The autoencoder generator methods \cite{erba2019real} can transfer the adversarial examples to follow the manifolds of the normal measurements (0\% detection accuracy). However, the size of the successful bad data is very smaller compared with the supreme attack and ConAML, which decreases the effect of the FDIA attack. In addition, we check the adversarial examples generated by \cite{erba2019real} and find that most of the generated examples (over 90\%) of still violate the physical constraints and will be removed by the residual-based detection in state estimation.

\begin{figure}[htbp]
\centerline{\includegraphics[width=1\linewidth]{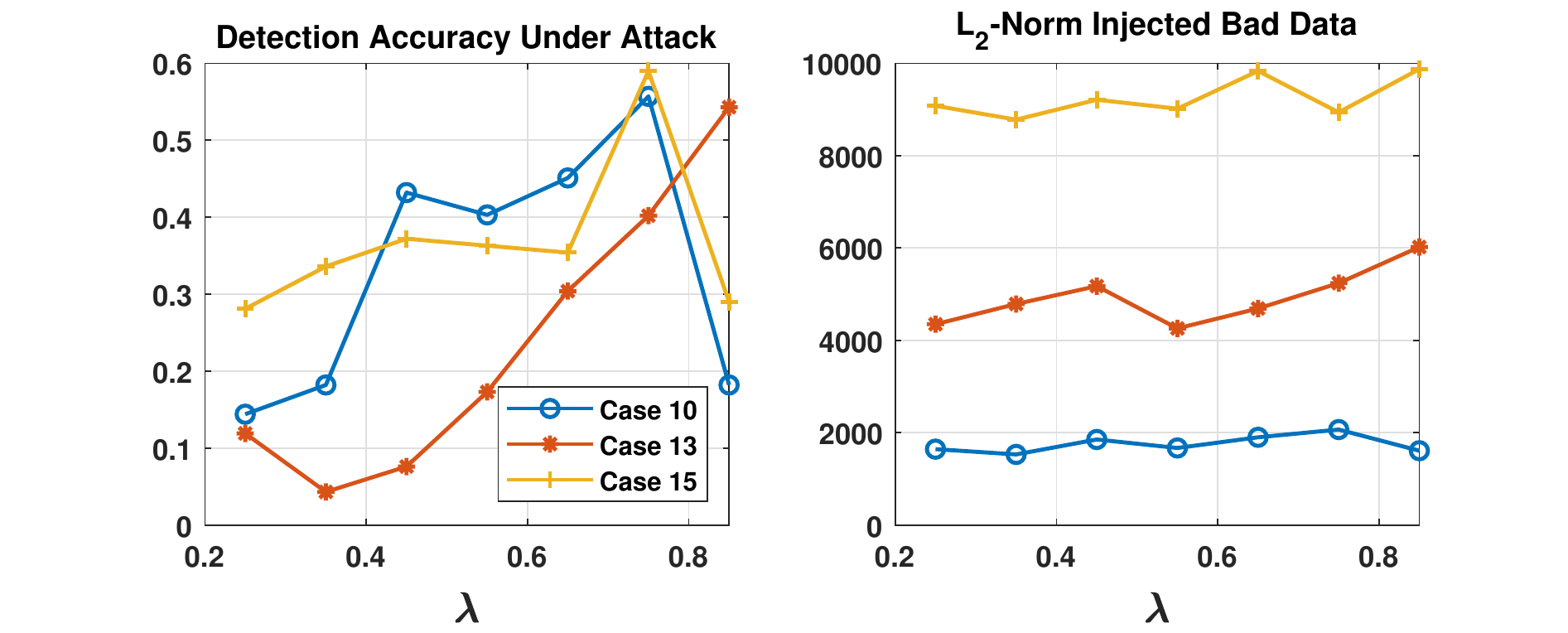}}
\caption{Performance of black-box attacks according to $\lambda$ with $step = 40$, $size = 20$.} 
\label{fig:powerLambda}
\end{figure}

As shown in Figure \ref{fig:powerLambda}, by comparing the evaluation results of different cases, we can learn that compromising more sensors cannot guarantee better performances in attack detection. This is due to the different physical constraints imposed by the system. However, with more compromised sensors, the attacker can usually obtain a larger size of the injected bad data.

\begin{figure}[htbp]
\centerline{\includegraphics[width=1\linewidth]{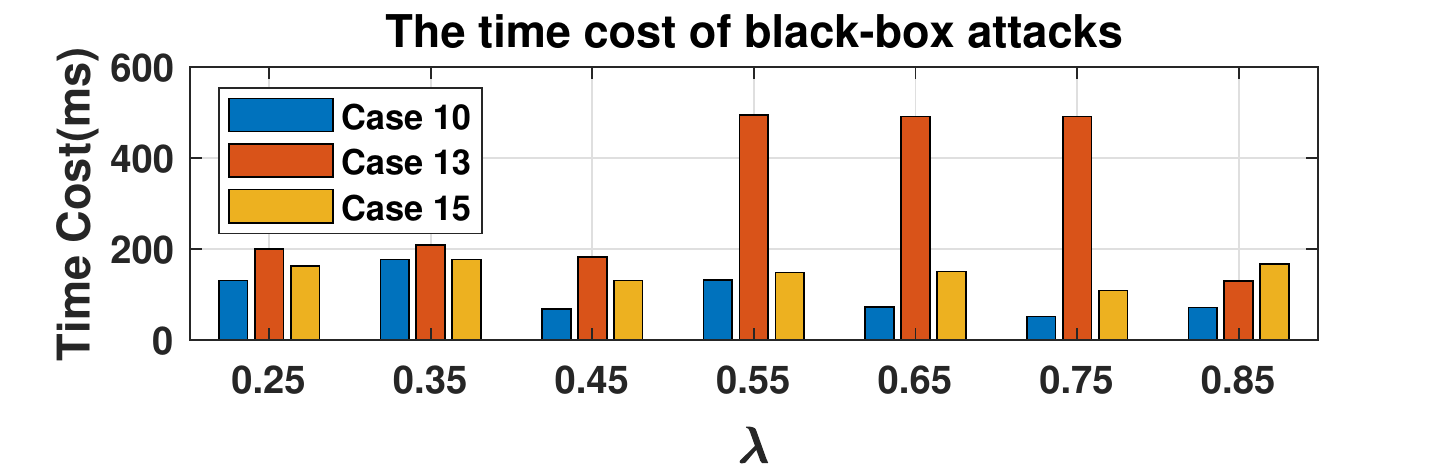}}
\caption{Time cost of black-box attacks according to $\lambda$ with $step = 40$, $size = 20$.} 
\label{fig:powerTimeLambda}
\end{figure}

In our experiments, the time cost of gray-box2 and black-box is much higher than other attack scenarios due to the universal adversarial measurements algorithm, as shown in Figure \ref{fig:powerTimeLambda}. However, the time cost is still efficient for many CPS applications in practice. For example, the sampling period of the traditional SCADA system used in power systems is 2 to 4 seconds. In practical scenarios, the time cost also depends on the computational resource of the attacker. With the possible optimization and upgrade in software and hardware, the time cost can be further reduced.

\subsection{Case Study: Water Treatment System}

\subsubsection{Background: SWaT Dataset}

In this section, we study the linear inequality physical constraints based on the Secure Water Treatment (SWaT) proposed in \cite{goh2016dataset}. SWaT is a scaled-down system but with fully operational water treatment functions. The testbed has six main processes and consists of cyber control (PLCs) and physical components of the water treatment facility. The SWaT dataset, generated by the SWaT testbed, is a public dataset to investigate the cyber attacks on CPSs.  The raw dataset has 946,722 samples with each sample comprised of 51 attributes, including the measurements of 25 sensors and the states of 26 actuators. Each sample in the dataset was labeled with normal or attack. \cite{goh2016dataset} investigated four kinds of attacks based on the number of attack points and places. The detailed description of the SWaT dataset can be found in \cite{goh2016dataset} and \cite{swatDataset}.

The SWaT dataset is an important resource to study anomaly detection in CPSs. Inoue \etal used unsupervised machine learning, including Long Short-Term Memory (LSTM) and SVM, to perform anomaly detection based on the SWaT dataset \cite{inoue2017anomaly}. By comparison, Kravchik \etal employed Convolutional Neural Networks (CNN) and achieved a better false positive rate \cite{kravchik2018detecting}. In 2019, \cite{feng2019systematic} proposed a data-driven framework to derive invariant rules for anomaly detection for CPS and utilized SWaT to evaluate their approach. Other research related to the SWaT dataset can be found in  \cite{chen2018learning, feng2017deep, ahmed2018noise}.

\begin{table}[htbp]
\caption{SWaT Analog Components}
\begin{center}
\begin{tabular}{|c|c|c|}
\hline
\textbf{Symbol} & \textbf{Description} & \textbf{Unit}\\
\hline
\textbf{LIT} & Level Indication Transmitter & $mm$ \\
\hline
\textbf{FIT} & Flow Indication Transmitter & $m^{3}/hr$ \\
\hline
\textbf{AIT} & Analyzer Indication Transmitter & $uS/cm$ \\
\hline
\textbf{PIT} & Pressure Indication Transmitter & $kPa$ \\
\hline
\textbf{DPIT} & Differential Pressure Ind Transmitter & $kPa$ \\
\hline

\end{tabular}
\end{center}
\label{table:swatfeature}
\end{table}

As shown in Table \ref{table:swatfeature}, the SWaT dataset includes the measurements from five kinds of analog components (25 sensors in total) whose measurements are used as the input features in previous anomaly detection ML models. Our experiments aims to demonstrate that the ML models used for anomaly detection are vulnerable to adversarial attacks. However, due to the physical properties of the SWaT testbed, the sensor's measurements are not independent but with linear inequality constraints. 

In our experiment, we consider the scenario that the attacker compromises the \textbf{FIT} components to inject bad adversarial water flow measurements. We examined the SWaT testbed structure and find out that there are apparent linear inequality constraints among the \textbf{FIT} measurements. The linear inequality constraints of the seven \textbf{FIT} measurements in the dataset are defined by the structure of the water pipelines and the placement of the sensors, as shown in equation \ref{eq:waterConst1}, where $\epsilon_{1}$ and $\epsilon_{2}$ are two constants of the system's noise tolerance. We checked the SWaT dataset and observed that all the normal examples in the dataset meet the constraints. We also contacted the managers of the SWaT testbed and verified our find. 

\begin{subequations}
	\begin{align}
	& \textbf{FIT301} \leq \textbf{FIT201}\\
	&\left \| \textbf{FIT401} - \textbf{FIT501}\right \|  \leq \epsilon_{1}  \\
	&\left \| (\textbf{FIT502} + \textbf{FIT503}) - (\textbf{FIT501} + \textbf{FIT504})\right \|  \leq \epsilon_{2} 
	\end{align}
\label{eq:waterConst1}
\end{subequations}

In our experiment, we show that the attacker can construct adversarial \textbf{FIT} measurements that can bypass the ML anomaly detection proposed in previous research. Meanwhile, our adversarial measurements will also follow the same linear inequality constraints to avoid being noticed by the system operator. 

\subsubsection{Experimental Design and Evaluation}

Similar to the power system study case, we generate two training datasets for the defender's model $f_{\theta}$ and the attacker's model $f'_{\theta'}$ respectively by poisoning the normal measurements with Gaussian noise. The ML models are trained to distinguish the normal measurement data and the poisoned measurements (anomaly). In our experiment, the overall classification accuracy of $f_{\theta}$ and $f'_{\theta'}$ is 97.2\% and 96.7\% respectively. After that, we consider the scenarios that there were 2, 5, and 7 \textbf{FIT} measurements compromised by the attacker and generate the related test datasets. The goal of the attacker is to generate the adversarial \textbf{FIT} measurements with the constraints defined by equation (\ref{eq:waterConst1}) so that the poisoned measurements can be classified as `normal' by $f_{\theta}$. A more detailed introduction of the experiment design and implementation, including the specific compromised measurements and the corresponding constraint matrix $\Phi$, can be found in \textbf{Appendix \ref{app:waterCons}} and \textbf{\ref{app:waterExpr}}.

\begin{table}[htbp]
\caption{Evaluation Result Summary}
\begin{center}
\begin{tabular}{|c|c|c|c|c|}
\hline
\textbf{Attack} & \textbf{Case} & \textbf{Accu} & \textbf{$L_{2}$-Norm} & \textbf{Time (ms)}\\
\hline
\multirow{3}*{Supreme } & 2 & 0\% & 3.24  & 3.59 \\
\cline{2-5}
  ~  & 5 & 0\% & 2.85 & 6.58 \\
\cline{2-5}
 ~ & 7 & 0\% & 4.31 & 9.21 \\
 \hline
 \multirow{3}*{Erba \cite{erba2019real}} & 2 & 85.3\% & 0.176 & 4.72 \\
\cline{2-5}
  ~  & 5 & 86.1\% & 0.017 & 4.12 \\
\cline{2-5}
 ~ & 7 & 88.4\% & 0.184 & 6.18 \\
 \hline
 \multirow{3}*{white-box} & 2 & 0\% & 0.741 & 21.7 \\
\cline{2-5}
  ~  & 5 & 0\% & 0.75 & 24.8 \\
\cline{2-5}
 ~ & 7 & 2\% & 1.05 & 71.5 \\
 \hline
 \multirow{3}*{gray-box1} & 2 & 0\% & 0.522 & 21.4\\
\cline{2-5}
  ~  & 5 & 53.2\% & 0.466 & 51.0\\
\cline{2-5}
 ~ & 7 & 89.3\% & 0.835 & 136.4 \\
 \hline
 \multirow{3}*{gray-box2} & 2 & 1.0\% & 0.52 & 42.9\\
\cline{2-5}
  ~  & 5 & 1.2\% & 0.627 & 94.2\\
\cline{2-5}
 ~ & 7 & 1.3\% & 0.841 & 256.1\\
 \hline
  \multirow{3}*{black-box} & 2 & 1.3\% & 0.309 & 17.5 \\
\cline{2-5}
  ~  & 5 & 2.3\% & 0.340 & 111.7\\
\cline{2-5}
 ~ & 7 & 1.14\% & 0.411 & 451.8 \\
 \hline
\end{tabular}
\end{center}
\label{table:waterEva}
\end{table}

Table \ref{table:waterEva} summarizes the evaluation performances of different scenarios of ConAML attacks. From the table, we can learn that the ConAML framework can still effectively decrease the detection accuracy of the ML models, even for black-box attacks. Meanwhile, even the black-box attack achieves a better performance on both the detection accuracy and bad data size compared with the baseline \cite{erba2019real}. The size of the injected bad data of the ConAML attacks is smaller than the supreme attacker. We explain that this is due to the stringent constraints between the \textbf{FIT} measurements. Similar to the power system study case, a larger number of compromised sensors cannot produce a better performance in bypassing the detection. The reason for this result is that more compromised sensors will also have more complex constraints between their measurements. Meanwhile, more constraints will increase the computation overhead of the best effort search algorithms since there will be a `larger' constraint matrix.

\begin{figure}[htbp]
\centerline{\includegraphics[width=1\linewidth]{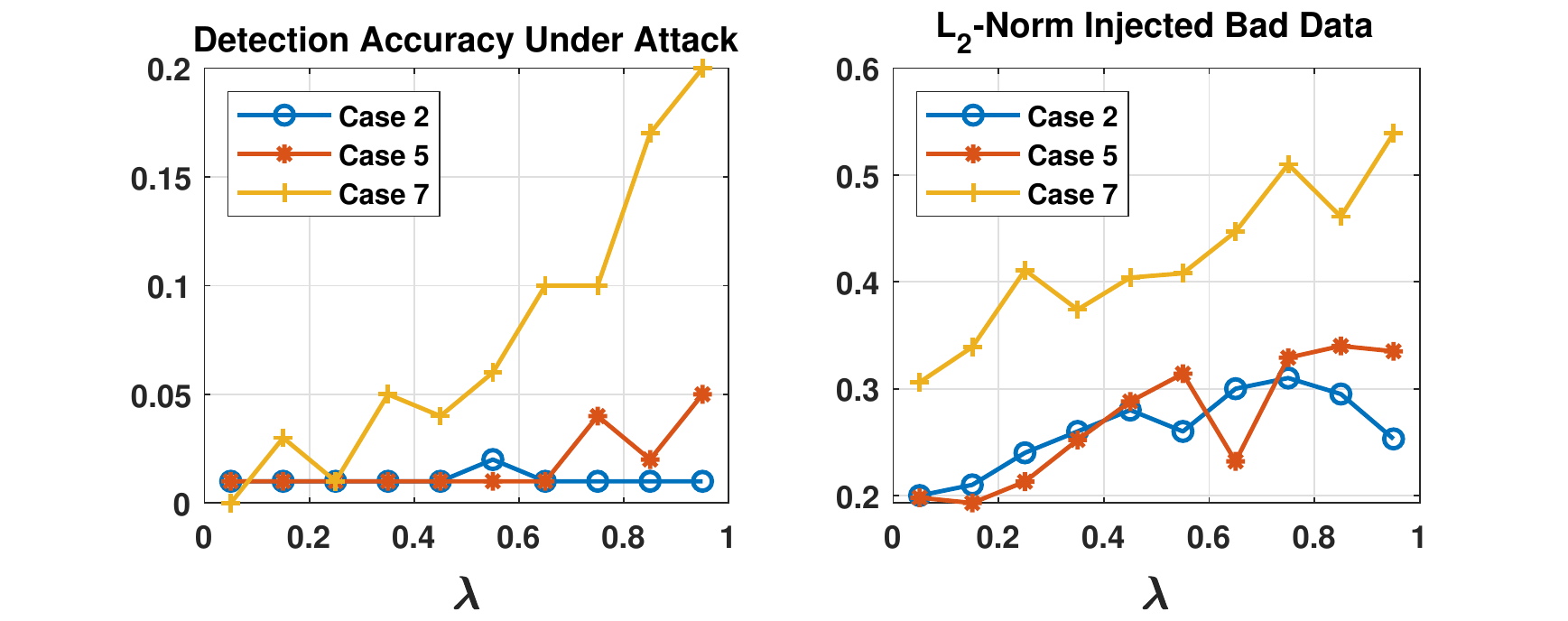}}
\caption{Performance of black-box attacks according to $\lambda$ with $step = 50$, $size = 0.06$.} 
\label{fig:waterLambda}
\end{figure}

Figure \ref{fig:waterLambda} demonstrated the trend of the detection accuracy and injected bad data size according to $\lambda$. From the figure, we can learn that, with the $\lambda$ increases, the probability of the adversarial examples being detected also increases. This matches the intuition that if an adversarial example can obtain higher successful attack probability with the sampling measurement set, its probability of evading detection will also increase. Meanwhile, a smaller injected data size is expected to make the adversarial examples look more `normal' to the detection model.
\section{Discussion and Future Work}
\label{sec:future}

As we mentioned in Section \ref{sec:intro}, in this paper, we mainly investigate the linear constraints of input measurements in CPSs and neural network-based ML algorithms. In the future, research on ConAML of nonlinear constraints and other general ML algorithms, such as SVM, KNN will be proposed. We encourage related communities to present different CPSs that require special constraints. 

As summarized in \cite{yuan2019adversarial}, defense mechanisms like adversarial re-training and adversarial detecting can increase the robustness of neural networks and are likely to mitigate ConAML attacks. However, most defenses in previous research target adversarial examples in computer vision tasks. In future work, we will study the state-of-the-art defense mechanisms in previous research and evaluate their performance with adversarial examples generated by ConAML. We will also investigate the defense mechanisms which take advantage of the properties of physical systems directly, such as the best deployment of sensors that will make the attackers' constraint more stringent.

\section{Conclusion}
\label{sec:conclusion}

The potential vulnerability of ML applications in CPSs need to be concerned. In this paper, we investigate the input constraints of AML algorithms in CPSs. We analyze the difference of adversarial examples between CPS and computational applications, like computer vision, and give the formal threat model of AML in CPS. We propose the best-effort search algorithms to effectively generate the adversarial examples that meet the linear constraints. Finally, as proofs of concept, we study the vulnerabilities of ML models used in FDIA in power grids and anomaly detection in water treatment systems. The evaluation results show that even with the constraints imposed by the physical systems, our approach can still effectively generate the adversarial examples that will significantly decrease the detection accuracy of the defender's ML models.

\bibliographystyle{plain}
\bibliography{reference}
\appendix

\section{Non-Linear Constraints}
\label{app:nonlinear}

Many other ML applications in the CPS domains, for instance, load forecasting in power and water systems, traffic forecasting in transportation systems, may have nonlinear constraints. The non-linear constraints can be very complex in various CPSs and cannot be covered in one study. In general, similar to linear constraints, the $k$ nonlinear constraints of the compromised measurements can be represented as equation (\ref{eq:non-linearCon}), where $\mu_{i}$ is a nonlinear function of $M_{C}$.

\begin{equation}
\left\{
             \begin{array}{lr}
             \mu_{0}( m_{c_{0}},  m_{c_{1}}, ...,  m_{c_{r-1}}) = 0 \\
            \mu_{1}( m_{c_{0}},  m_{c_{1}}, ...,  m_{c_{r-1}}) = 0 \\
              ...\\
             \mu_{k-1}( m_{c_{0}},  m_{c_{1}}, ...,  m_{c_{r-1}}) = 0 \\
             \end{array}
\right.
\label{eq:non-linearCon}
\end{equation}

We now investigate a special case of the nonlinear constraints. If there exists a subset of the compromised measurements, in which each measurement can be represented as an explicit function of the measurements in the complement set, the attacker will also be able to generate the perturbation accordingly. We use $P = \left [p_{0}, p_{1}, ..., p_{n-1} \right ]$ to denote the index vector of the former measurement set, and use $Q = \left [q_{0}, q_{1}, ..., q_{r-n-1} \right ]$ to denote the index vector of the complement set. We can then represent (\ref{eq:non-linearCon}) as (\ref{eq:non-linearSpe}), where $\Xi = \left [ \xi_{0},  \xi_{1}, ..., \xi_{n-1} \right ]$ is a vector of explicit functions.

\begin{equation}
\left\{
             \begin{array}{lr}
             m_{p_{0}} = \xi_{0}(m_{q_{0}}, m_{q_{1}}, ..., m_{q_{r-n-1}}) \\
             m_{p_{1}} = \xi_{1}(m_{q_{0}}, m_{q_{1}}, ..., m_{q_{r-n-1}}) \\
              ...\\
             m_{p_{n-1}} = \xi_{n-1}(m_{q_{0}}, m_{q_{1}}, ..., m_{q_{r-n-1}}) \\
             \end{array}
\right.
\label{eq:non-linearSpe}
\end{equation}

Apparently, the roles of $M_{Q}$ and $M_{P}$ in (\ref{eq:non-linearSpe}) are similar to the $M_{I}$ and $M_{D}$ in linear constraints correspondingly. Instead of a linear matrix, the function set $\Xi$ represents the dependency between $M_{P}$ and $M_{Q}$. The nonlinear constraints make properties such as Theorem 1 infeasible. To meet the constraints, the attacker needs to find the perturbation $\Delta_{Q}$ first and obtain $M^{\ast}_{Q}$ by adding it to $M_{Q}$. After that, the attacker can compute $M^{\ast}_{P} = \Xi(M^{\ast}_{Q})$ .

The above case of nonlinear constraints is special and may not be scalable to various practical applications. Although there are different types of nonlinear systems, they can be generalized using piece-wise linear constraints by setting proper ranges and breakpoints. We leave this as an open problem for future work.

\section{Proofs}
\label{app:proof}

\subsection{Theorem \ref{theorem:condition}}

\begin{proof}
If we replace $M_{C}^{\ast}$ in equation (\ref{eq:conOpt}d) with equation (\ref{eq:conOpt}b), we can get $\Phi_{k \times r}  M_{C}^{\ast} = \Phi_{k \times r} ( M_{C} + \Delta _{C}) = \Phi_{k \times r}  M_{C} + \Phi_{k \times r} \Delta _{C} = \tilde{\Phi}$. From equation (3c) we can learn that $\Phi_{k \times r}  M_{C} = \tilde{\Phi}$. Therefore, we have $\Phi_{k \times r} \Delta _{C} = 0$ and prove Theorem \ref{theorem:condition}.
\end{proof}

\subsection{Corollary \ref{corollary:combine}}

\begin{proof}
We have $\Phi_{k \times r} \Delta_{C'} = \Phi_{k \times r} \sum_{i=0}^{n} a_{i} \cdot \Delta_{C_{i}} = \sum_{i=0}^{n} a_{i} \cdot \Phi_{k \times r}\Delta_{C_{i}}$. Since $\Delta_{C_{i}}$ is a valid perturbation vector and $\Phi \Delta_{C_{i}} = 0$, we have $\Phi_{k \times r} \Delta_{C'} = 0$ and prove Corollary \ref{corollary:combine}.
\end{proof}

\subsection{Theorem \ref{theorem:alwaysMet}}

\begin{proof}

Due to the intrinsic property of the targeted system, equation (3c) is naturally met, which indicates that there is always a solution for the nonhomogeneous linear equations $\Phi_{k \times r}  X = \tilde{\Phi}$. Accordingly, we have $Rank(\Phi_{k \times r}) \leq r$. Moreover, if $Rank(\Phi_{k \times r}) = r$, there will be one unique solution for equation (\ref{eq:conOpt}c), which means the measurements of compromised sensors are constant. The constant measurements are contradictory to the purpose of deploying CPSs. In practical scenarios, $M$ is changing over time, so that $Rank(\Phi_{k \times r}) < r$ and the homogeneous linear equation $\Phi_{k \times r} X = 0$ will have infinite solutions. Therefore, the attacker can always build a valid adversarial example that meets the constraints.

\end{proof}

\section{Power System Case Study}
\label{app:power}
\subsection{State Estimation and FDIA}
\label{app:fdia}

We give the mathematical description of state estimation and how a false data injection attack (FDIA) can be launched. To be clear, we will employ the widely used notations in related research publications to denote the variables; the corresponding explanation will also be given to avoid confusion.

In general, the AC power flow measurement state estimation model can be represented as follow:

\begin{equation}
\textbf{z} = \textbf{h}(\textbf{x}) + \textbf{e}
\label{eq:acModel}
\end{equation}

where $\textbf{h}$ is a function of $\textbf{x}$, $\textbf{x}$ is the state variables, $\textbf{z}$ is the measurements, and $\textbf{e}$ is the measurement errors. The task of state estimation is to find an estimated \textbf{$\hat{\textbf{x}}$} that best fits $\textbf{z}$ of (\ref{eq:acModel}). In practical application, a DC measurement model is also used to decrease the process time and (\ref{eq:acModel}) can then be represented as follow:

\begin{equation}
\textbf{z} = \textbf{H}\textbf{x} + \textbf{e}
\end{equation}
where \textbf{$\textbf{H}_{m \times n}$} is a matrix that determined by the topology, physical parameters and configurations of the power grid.

Typically, if a weighted least squares estimation scheme is used, the system state variable vector \textbf{$\hat{\textbf{x}}$} can be obtained through (\ref{eq:wlse}):

\begin{equation}
\textbf{$\hat{\textbf{x}}$} = (\textbf{H}^{T} \textbf{W} \textbf{H})^{-1} \textbf{H}^{T} \textbf{W} \textbf{z}
\label{eq:wlse}
\end{equation}
where \textbf{$\textbf{W}$} is the covariance matrix of the variances of meter errors.

Due to possible meter instability and cyber attacks, bad measurements may be introduced to the measurement vector $\textbf{z}$. To solve this, various bad measurement detection methods are proposed \cite{monticelli2012state}. One commonly used detection approach is to calculate the measurement residual between the raw measurement $\textbf{z}$ and derived measurements $\textbf{H} \textbf{$\hat{\textbf{x}}$}$. If the $L_{2}$-norm $\left \| \textbf{z} - \textbf{H} \textbf{$\hat{\textbf{x}}$} \right \| > \tau$, where $\tau$ is a threshold selected according to the false alarm rate, the measurement $\textbf{z}$ will be considered as a bad measurement.

The above detection method contains non-linear computation ($L_{2}$-Norm), however, research has shown that a false measurement vector follows linear equality constraints can be used to pollute the normal measurements without being detected. In 2009, Liu \etal proposed the false data injection attack (FDIA) that can bypass the detection scheme described above and pollute the result of state estimation \cite{liu2009false}. FDIA assumes that the attacker knows the topology and configuration information $\textbf{H}$ of the power system. Let $\textbf{z}_{a} = \textbf{z} + \textbf{a}$ denote the compromised measurement vector that is observed by the state estimation, where 
$\textbf{a}$ is the malicious data added by the attacker. Thereafter, let $\textbf{$\hat{\textbf{x}}$}_{bad} = \textbf{$\hat{\textbf{x}}$} + \textbf{c}$ denote the polluted state that is estimated by $\textbf{z}_\textbf{a}$, where $\textbf{c}$ represents the estimation error brought by the attack. Liu \etal demonstrated that, as long as the attacker builds the injection vector $\textbf{a} = \textbf{H} \textbf{c}$, the polluted measurements $\textbf{z}_\textbf{a}$ will not be detected by the measurement residual scheme.

\begin{proof}
If the original measurements $\textbf{z}$ can pass the detection, the residual $\left \| \textbf{z} - \textbf{H} \textbf{$\hat{\textbf{x}}$} \right \| \leq \tau$. Through (\ref{eq:ahc}) from \cite{liu2009false}, we learn that the measurement residual will be the same when $\textbf{a} = \textbf{H} \textbf{c}$. Therefore, the crafted measurements from the attacker will not be detected.
\end{proof}

\begin{subequations}
\begin{align}
\left \| \textbf{z}_{a} - \textbf{H} \textbf{$\hat{\textbf{x}}$}_{bad} \right \| \;\; &= \;\; \left \| \textbf{z} + a - \textbf{H}(\textbf{$\hat{\textbf{x}}$} + \textbf{c}) \right \| \\
 \;\; &= \;\; \left \| \textbf{z} - \textbf{H}\textbf{$\hat{\textbf{x}}$} + (\textbf{a} - \textbf{H}\textbf{c}) \right \| \\
 \;\; &= \;\; \left \| \textbf{z} - \textbf{H}\textbf{$\hat{\textbf{x}}$} \right \| \leq \tau
\end{align}
\label{eq:ahc}
\end{subequations}

Besides, \cite{liu2009false} also provided the approach to effectively find vector $a$ that will meet the attack requirement. Let $\textbf{P} = \textbf{H}(\textbf{H}^{T}\textbf{H})^{-1}\textbf{H}^{T}$ and matrix $\textbf{B} = \textbf{P} - \textbf{I}$. In order to have $\textbf{a} = \textbf{H}\textbf{c}$, $\textbf{a}$ needs to be a solution of the homogeneous equation $\textbf{B}\textbf{X} = \textbf{0}$, as shown in (\ref{eq:Ba0}).

\begin{subequations}
\begin{align}
\textbf{a} = \textbf{H}\textbf{c} \;\; &\Leftrightarrow\;\; \textbf{P}\textbf{a} = \textbf{P}\textbf{H}\textbf{c} \Leftrightarrow \textbf{P}\textbf{a} = \textbf{H}\textbf{c} \Leftrightarrow \textbf{P}\textbf{a} = \textbf{a} \\
\;\; &\Leftrightarrow\;\; \textbf{P}\textbf{a} - \textbf{a} = \textbf{0} \Leftrightarrow  (\textbf{P}-\textbf{I})\textbf{a} = \textbf{0} \\
\;\; &\Leftrightarrow\;\; \textbf{B}\textbf{a} = \textbf{0}
\end{align}
\label{eq:Ba0}
\end{subequations}

Another problem of generating $\textbf{a}$ is when will (\ref{eq:Ba0}c) have a solution. Liu \etal prove that, suppose the attacker compromises $k$ meters, as long as $k > m-n$, there always exists non-zero attack vector $\textbf{a} = \textbf{H}\textbf{c}$. We refer the readers to \cite{liu2009false} for the detailed proof.

\subsection{Experiment Implementation}
\label{app:powerExper}

\begin{figure}[htbp]
\centerline{\includegraphics[width=0.98\linewidth]{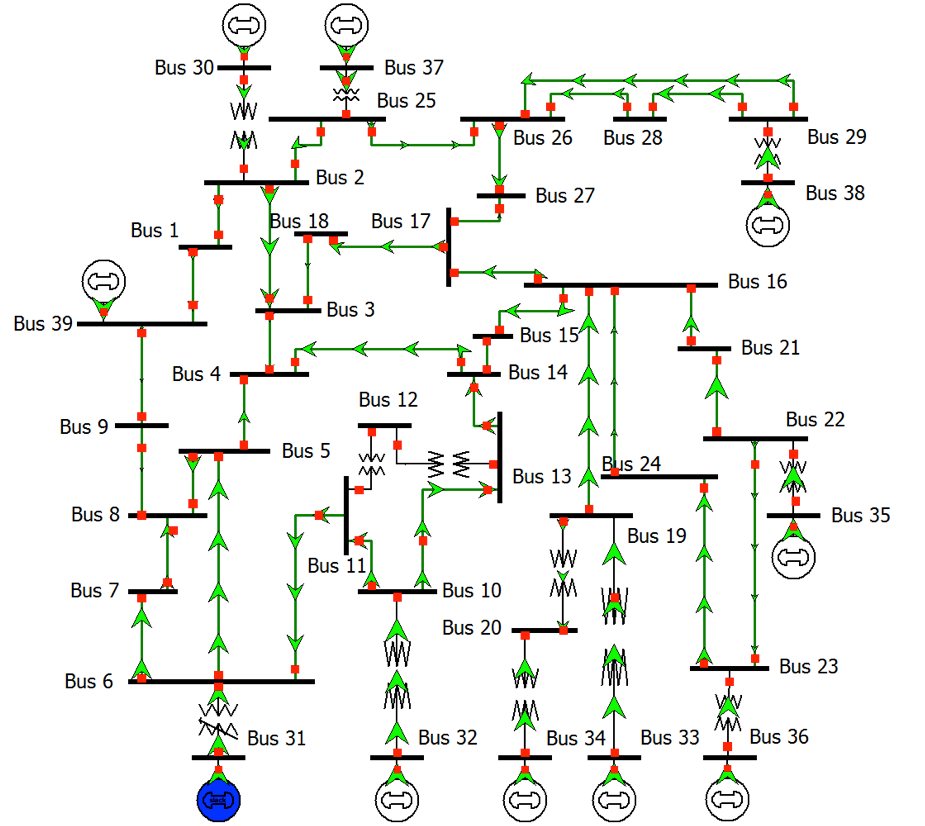}}
\caption{IEEE 39-Bus System \cite{athay1979practical} \cite{39busfigure}.}
\label{fig:39bus}
\end{figure}

The structure of the IEEE 39-bus system is shown in Figure \ref{fig:39bus}. We utilize the MATPOWER \cite{zimmerman2010matpower} library to derive the $\textbf{H}$ matrix of the system and simulate the power flow measurement data. We also implement the FDIA using MATLAB to generate false measurements. Both the power flow measurements and false measurements follow Gaussian distributions. We make two datasets for the defender and the attacker respectively. For each dataset, there are around 25,000 records with half records are polluted with FDIA. We label the normal measurements as 0 and false measurements as 1 and use one-hot encoding for the labels. 

We investigate the scenarios that there are 10, 13, and 15 measurements being compromised by the attacker, with the randomly generated compromised index vector $C$ and corresponding constraint matrix $\Phi$ ($\textbf{B}_{C}$ in (\ref{eq:Ba0})). We generate 1,000 false measurement vectors in each test datasets.

After that, we train two deep learning models based on the training datasets accordingly, with 75\% records in the dataset used for training and 25\% for testing. We use simple fully connected neural networks as the ML models and the model structures are shown in Table \ref{table:FDIAModel}. Both the models are trained with a 0.0001 leaning rate, 512 batch size, a mean squared error loss function, and a Stochastic Gradient Descent (SGD) optimizer. The deep learning models are implemented using Tensorflow and the Keras library and are trained on a Windows 10 machine with an Intel i7 CPU. The training process is around one minutes for each model. 

\begin{table}[htbp]
\caption{Model Structure - FDIA}
\begin{center}
\begin{tabular}{|c|c|c|}
\hline
Layer & $f$ & $f'$ \\
\hline
0 & 46 Input & 46 Input \\
\hline
1 & 32 Dense ReLU & 30 Dense ReLU \\
\hline
2 & 48 Dense ReLU & 40 Dense ReLU \\
\hline
3 & 56 Dense ReLU & 30 Dense ReLU \\
\hline
4 & 48 Dense ReLU & Dropout 0.25 \\
\hline
5 & 32 Dense ReLU & 20 Dense ReLU \\
\hline
6 & Dropout 0.25 &  Dropout 0.25 \\
\hline
7 & 16 Dense ReLU & 2 Dense Softmax \\
\hline
8 & Dropout 0.25  &  - \\
\hline
9 & 2 Dense Softmax &  - \\

\hline
\end{tabular}
\end{center}
\label{table:FDIAModel}
\end{table}

\section{Water Treatment Case Study}
\label{app:water}
\subsection{SWaT Measurement Constraints}
\label{app:waterCons}

We examined the user manual of the SWaT system and check the structure of the water pipelines. We found some \textbf{FIT} measurements in SWaT should always follow inequality constraints when the whole system is working steadily. Based on the component names described in \cite{goh2016dataset}, the constraints can be represented as (\ref{eq:waterConst}), where $\epsilon_{1}$ and $\epsilon_{2}$ are the allowed measurement errors. We utilized the double value of the maximum difference of the corresponding measurements in the SWaT dataset to estimate $\epsilon_{1}$ and $\epsilon_{2}$, and we had $\epsilon_{1} = 0.0403$ and $\epsilon_{2} = 0.153$. 

\begin{subequations}
	\begin{align}
	& \textbf{FIT301} \leq \textbf{FIT201}\\
	&\left \| \textbf{FIT401} - \textbf{FIT501}\right \|  \leq \epsilon_{1}  \\
	&\left \| (\textbf{FIT502} + \textbf{FIT503}) - (\textbf{FIT501} + \textbf{FIT504})\right \|  \leq \epsilon_{2} 
	\end{align}
\label{eq:waterConst}
\end{subequations}

Based on (\ref{equ:linearIneq}), we can represent (\ref{eq:waterConst}) as follow. And $M_{C}$ is the vector of measurements of \textbf{FIT201}, \textbf{FIT301}, \textbf{FIT401}, \textbf{FIT501}, \textbf{FIT502}, \textbf{FIT503} and \textbf{FIT504} accordingly.

\[
\Phi_{5 \times 7} = 
\begin{bmatrix}
-1 & 1 &  0 &  0 &  0 &  0 &  0\\ 
 0 & 0 &  1 & -1 &  0 &  0 &  0\\ 
 0 & 0 & -1 &  1 &  0 &  0 &  0\\ 
 0 & 0 &  0 & -1 &  1 &  1 & -1\\ 
 0 & 0 &  0 &  1 & -1 & -1 &  1
\end{bmatrix}
\tilde{\Phi} = 
\begin{bmatrix}
0\\
0.0403\\
0.0403\\
0.153\\
0.153
\end{bmatrix}
\]

We consider three scenarios that there are 2, 5, and 7 \textbf{FIT} measurements being compromised by the attacker, and the compromised sensors are $\left \{ \textbf{FIT201}, \textbf{FIT301} \right \}$, $ \{ \textbf{FIT401}, \textbf{FIT501}, \textbf{FIT502},$  
$\textbf{FIT503}, \textbf{FIT504} \}$, and all the seven \textbf{FIT} sensors respectively. The constraint matrix of each scenario can be derived from the corresponding rows of the $\Phi_{5 \times 7}$ matrix.

\subsection{Experimental Implementation}
\label{app:waterExpr}

In the Swat dataset, we extracted the normal records which were sampled when the whole system was working steadily. We also removed all the actuators' features. Here, we denote the extracted records as $D_{e}$. After that, we randomly picked out three test datasets from $D_{e}$ as the with each test dataset contains 1000 records. We added Gaussian noise to the compromised measurements of records in all test datasets. We checked the polluted record every time when a noise vector was added to ensure all the records in test datasets meet the linear inequality constraints. Here, we denote the rest records of $D_{e}$ as $D_{train}$ which contains 120,093 records with each record having 25 features in our implementation. We randomly and equally split $D_{train}$ into $D_{defender}$ and $D_{attacker}$ for the defender and attacker respectively and pollute half records with normally-distributed random noise in $D_{train}$ and $D_{defender}$. The polluted records in $D_{defend}$ and $D_{attacker}$ are labeled with 1 and the rest with 0. We allow the records in $D_{train}$ and $D_{defend}$ with label 1 to violate the constraints since the ML models are also expected to detect the obviously anomalous measurements. 

We utilize $D_{defend}$ and $D_{attack}$ to train the ML models $f_{\theta}$ and $f'_{\theta'}$ for the defender and attacker respectively. Again, 75\% records in the both datasets were used for training the 25\% records for testing. Similar to the FDIA experiment, we utilize fully connected neural networks and the structures are shown in Table \ref{table:swatModel}. Through parameter tuning, model $f_{\theta}$ and $f'_{\theta'}$ achieves 97.2\% and 96.7\% accuracy respectively.

\begin{table}[htbp]
\caption{Model Structure - Water Treatment}
\begin{center}
\begin{tabular}{|c|c|c|}
\hline
Layer & $f$ & $f'$ \\
\hline
0 & 25 Input & 25 Input \\
\hline
1 & 20 Dense ReLU & 24 Dense ReLU \\
\hline
2 & 40 Dense ReLU & 32 Dense ReLU \\
\hline
3 & 30 Dense ReLU & 32 Dense ReLU \\
\hline
4 & Dropout 0.25  & 16 Dense ReLU \\
\hline
5 & 20 Dense ReLU & 2 Dense Softmax \\
\hline
6 & Dropout 0.25  &  - \\
\hline
7 & 2 Dense Softmax & - \\

\hline
\end{tabular}
\end{center}
\label{table:swatModel}
\end{table}

\clearpage

\end{document}